\documentclass[3p,11pt,authoryear]{elsarticle}

\usepackage{amssymb}
\usepackage{amsmath}
\usepackage{mathrsfs}
\usepackage{rotating}
\usepackage{graphicx}
\usepackage{epstopdf}
\usepackage{longtable,lscape}

\newcommand{\fn}[1]{\footnote{\scriptsize{#1}}} 

\newcommand{\rres}{r_\mathrm{res}}
\newcommand{\Omegares}{n_\mathrm{res}}
\newcommand{\Eqn}[1]{Eq{#1}.}  
\newcommand{\Fig}[1]{Fig{#1}.}  
\newcommand{\ud}{\mathrm{d}}  
\newcommand{\Cassit}{\textit{Cassini}}  
\newcommand{\Voyit}{\textit{Voyager}}  

\usepackage{floatrow}

\journal{Icarus}

\begin{document} 

\begin{frontmatter}
\title{Mapping spiral waves and other radial features in Saturn's rings} 

\author{Matthew~S.~Tiscareno$^{1}$\fn{Formerly of the Center for Radiophysics and Space Research, Cornell University, Ithaca, NY 14853, USA.} and Brent~E.~Harris$^2$}

\address{$^1$Carl Sagan Center for the Study of Life in the Universe, SETI Institute, \\189 Bernardo Avenue \#200, Mountain View, CA 94043, USA.\\$^2$Department of Earth and Space Sciences, University of California, Los Angeles, CA 90095, USA.}

\begin{abstract}
We have analyzed the highest-quality images to be obtained by \Cassit{} of Saturn's main rings after the Saturn Orbit Insertion (SOI) and before the Ring Grazing Orbits (RGO) and Grand Finale (GF).  These images are comparable to those of SOI in fidelity, though not in nominal resolution, due to their high signal-to-noise.  We have systematically searched for radial structure in these images by reducing them to a single dimension (distance from Saturn's center) and using the continuous wavelet transform technique.  We discuss the resonant theory of spiral waves and discuss the proper method for deriving the local surface density from the wavelet signature of a spiral wave.  We present 1) individual features of interest found in our data, including several classes of waves that have not previously been reported; 2) a radial profile of surface density in Saturn's rings, which is more definitive for the A~ring than any previously presented and which corrects some errors in previous profiles; and 3) an atlas of resonant features that indicates whether each feature is or is not expressed in the rings and that is organized graphically by resonance strength. 
\end{abstract}


\end{frontmatter}

\section{Introduction}

The rings of Saturn, in addition to being among the most beautiful and enticing structures in the solar system, constitute a dynamic system with a great deal of internal structure.  The vast majority of that structure is too fine to be seen in the global-scale views of the rings that are most often seen by the public, while close-up views are often presented without sufficient context to enable viewers to comprehend how the whole is composed of the parts.  The purpose of this work is to take a first step towards an ``atlas'' of Saturn's rings, which presents a global view yet with enough detail to see the fine structure, and which also uses several tools to explicate the fine structure. 

Nearly all the fine structure in Saturn's rings is either azimuthally symmetric or in the form of a tightly-wound spiral; moreover, the spiral structures can be understood as phased wavefronts passing through a spatial frequency profile that is azimuthally symmetric.  Therefore, a radial profile of the ring is sufficient to account for the basic structure of the ring system.  Azimuthally compact structures do exist, such as spokes \citep{Mitchell13} and propellers \citep{Propellers06,Propellers08,Giantprops10,Sremcevic07}, but these may be considered as objects moving across a ``landscape,'' while the landscape itself is azimuthally symmetric and is the subject of this work. 

This work focuses on the ring's structure as seen in images obtained by the Imaging Science Subsystem (ISS) on board the \Cassit{} spacecraft.  The ISS~images generally have substantially lower resolution than occultation scans from the VIMS, UVIS, and RSS instruments.  However, occultations are one-dimensional scans, and as such often have difficulty distinguishing actual radial ring structure from local ring microstructure \citep[e.g.,][]{SN90}.  On the other hand, because ISS~images are two-dimensional, averages over the azimuthal dimension can reveal very subtle radial structure, especially when exposures are set to take advantage of the camera's sensitivity and obtain optimal signal-to-noise.  Therefore, ISS~images are potentially the best instrument for detecting a certain class of subtle radial features, as long as their radial extent is not much smaller than the nominal resolution of the images. 

The maps and atlases of ring structure presented in this work are innovative and (we hope) useful, but are not yet definitive.  They should be improved in the future by incorporating occultation data from the \Cassit{} RSS, UVIS, and~VIMS instruments.  They should also be improved by incorporating \Cassit{}~ISS data that was recently obtained during the Ring Grazing Orbits (RGOs) and the Grand Finale (GF) in 2016 and~2017, during which the spacecraft passed repeatedly very close to the rings and obtained a radially complete series of images at much better resolution than the images presented in this work.  The RGO/GF images will be presented in future work. 

Section~\ref{Theory} reviews relevant portions of the theory of spiral waves, which is at the heart of much of this work.  Sections~\ref{Observations} and~\ref{Methods} describe the methods by which the data we present were obtained and processed.  Section~\ref{Results} describes our results, including individual features of interest found in our data, a radial profile of surface density in Saturn's rings, and an atlas of resonant features that indicates whether each feature is or is not expressed in the rings and that is organized graphically by resonance strength.  

\section{Theory \label{Theory}}

Spiral waves occur at locations in the rings where the forcing frequency from the moon forms a whole-number ratio (the ``resonance label'') with the orbital frequency\fn{It is actually the ring particle's radial frequency $\kappa$ or vertical frequency $\nu$ that respectively resonates with the forcing frequency to create a SDW or a SBW.  However, both $\kappa$ and $\nu$ differ from the mean motion $n$ by a small factor of order $J_2$, and it is sufficient to think in terms of $n$ when computing the resonance label.} of ring particles.  \textit{Lindblad resonances} excite the eccentricities of ring particles, leading to spiral density waves (SDWs).  \textit{Vertical resonances} excite the inclinations of ring particles, leading to spiral bending waves (SBWs). 

\subsection{Wave dispersion and background surface density \label{Dispersion}}
The linear theory of spiral waves in disks has been described in several classic publications 
\citep[see review by][]{SchmidtChapter09}. 
For our purposes, we only need quote the relation between a wave's spatial frequency $k = 2 \pi / \lambda$ (for wavelength $\lambda$) and the distance between any radial point $r$ (measured from Saturn's center within the ring plane) and the resonance point $\rres$, at which the wave is generated: 
\begin{equation}
\label{WavenumEqn}
k(r) = \frac{\mathscr{D}}{2 \pi G \sigma_0 \rres} (r-\rres), 
\end{equation}
\noindent where $G$ is Newton's constant, $\sigma_0$ is the background surface density at radius $r$, and the factor $\mathscr{D}$ is given by
\begin{equation}
\label{ScriptyD}
\mathscr{D} = \begin{cases}
3(m-1)\Omegares^2, & m > 1 \\ 
\frac{21}{2} J_2 \frac{R_\mathrm{Sat}^2}{\rres^2} \Omegares^2, \hspace{1cm} & m = 1 
\end{cases} 
\end{equation}
\noindent where $\Omegares$ is the orbital frequency (or ``mean motion'') at radial location $\rres$, $J_2$ is the quadrupole gravity harmonic of Saturn, $R_\mathrm{Sat} = $~60,330~km is the radius\fn{Note that $R_\mathrm{Sat}$ is not subject to refinement by improved understanding of Saturn's atmosphere, as it is only a conventional value by which $J_2$ is normalized.} of Saturn, and $m$ is the azimuthal parameter\fn{One can determine $m$ from the identity of the resonance, as a first-order resonance is labeled $m$:($m$-1), a second-order resonance is labeled ($m$+1):($m$-1), a third order resonance is labeled ($m$+2):($m$-1), and so on.} of the resonance.

Nearly all spiral waves discussed in this paper are described by the upper branch of \Eqn{}~\ref{ScriptyD}.  First-order resonances (for which the two numbers in the ratio differ by one) are the strongest for any given moon, and moons as small as Pan and Atlas (but not Daphnis) raise observed SDWs at their first-order resonances.  Higher-order resonances can also raise observable waves, if the perturbing moon is relatively massive and has a significant eccentricity (for SDWs) or inclination (for SBWs).  Two factors cause SBWs to be much rarer than SDWs, namely that vertical resonances cannot be first-order, and that inclinations for moons in the Saturn system are generally small.  Saturn's small, close-in ``ring moons'' Pan, Atlas, Prometheus, Pandora, Janus, and Epimetheus produce abundant SDWs in the main rings.  More massive and more distant moons, such as Mimas, produce higher-order SDWs and SBWs, some of which are more powerful than the first-order waves driven by the smaller moons. 

Resonances for which $m=1$, described by the lower branch of \Eqn{}~\ref{ScriptyD}, are those for which the forcing frequency from the moon resonates not with the ring particle's orbital frequency but with its precession frequency \citep{Cuzzi81,RL88,Iapetuswave13}.  Because precession frequencies are much slower than orbital frequencies, the moons that raise $m=1$ waves are much farther from the planet than those that raise $m > 1$ waves; in the case of Saturn's rings, observed $m=1$ waves are raised by Titan, Hyperion, and Iapetus. 

\begin{figure}[!t]
\begin{center}
\includegraphics[height=7cm]{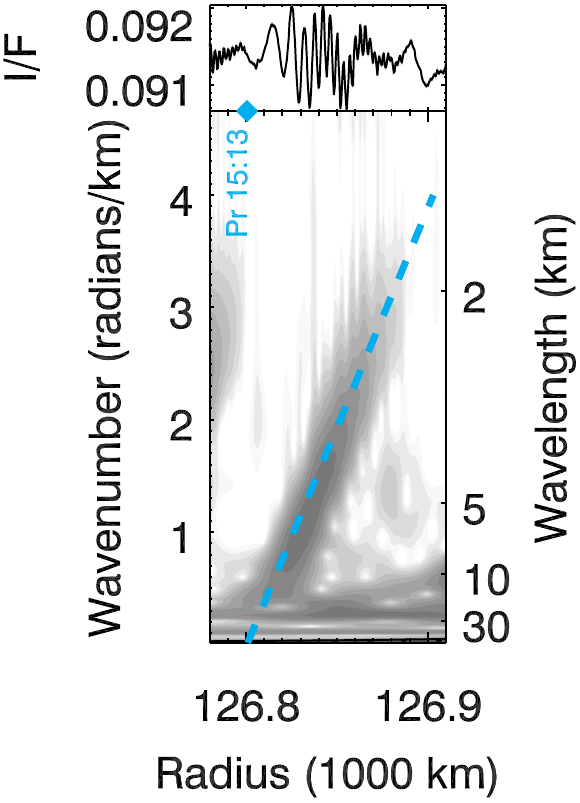}
\hspace{2cm}
\includegraphics[height=7cm]{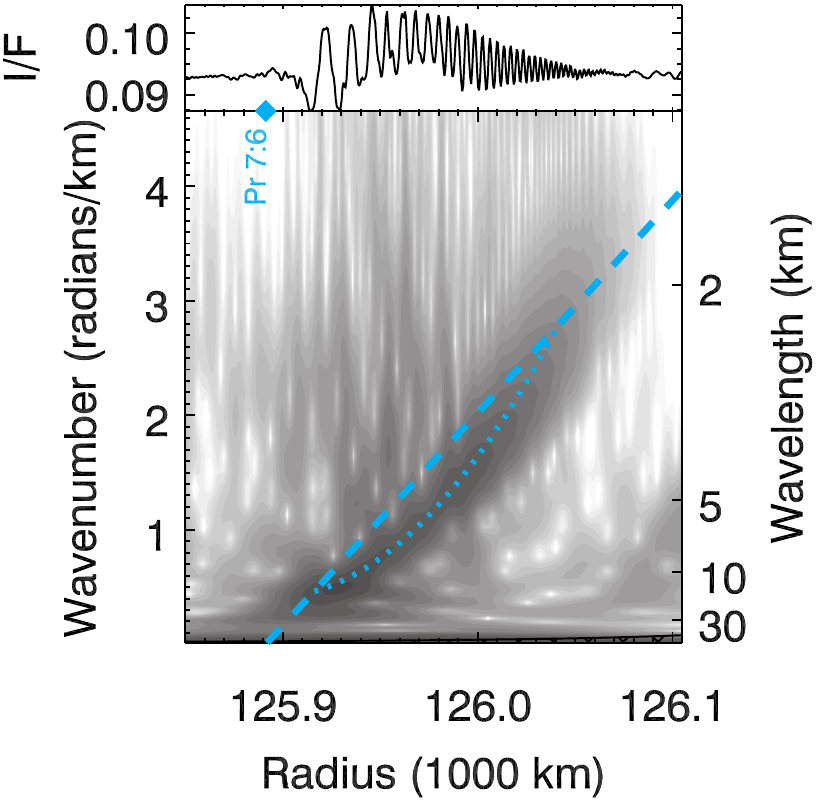}
\caption{Wavelet transforms are spatially-resolved power spectra in which darker gray shades indicate greater power in a particular spatial wavenumber $k$ at a particular radius $r$ \citep[for details see][]{soirings}.  The right-hand axis of each plot shows the wavelength $\lambda = 2 \pi / k$.  Across the top is the image radial scan (see Section~\ref{Observations}) of which the wavelet transform has been taken.  On the line between the radial scan and the wavelet transform are diamonds showing resonance locations with text giving corresponding resonance labels, both color-coded by the perturbing moon (in this case, cyan for Prometheus).  Dashed lines on the wavelet transform with the same color-coding indicate predicted slopes from \Eqn{}~\ref{WavenumEqn} with a constant value of $\sigma_0$.  (\textit{left}) The trace of a linear SDW is a slanted straight line, coincident with the predicted slope.  Depicted is the Prometheus~15:13 SDW from image N1560310964, with $\sigma_0 = 35$~g~cm$^{-2}$.  (\textit{right}) The trace of a non-linear SDW is concave-up (following the freehand-drawn dotted line), appearing to hang downward from the predicted slope.  This is because inter-particle effects, which are not modeled by linear theory, concentrate mass in the central regions of non-linear waves, effectively raising $\sigma_0$ in such regions.  Also seen in this figure is a larger amount of noise surrounding the main feature, which is due to non-sinusoidal morphology in the wave.  Depicted is the Prometheus~7:6 SDW from image N1560311082, with $\sigma_0 = 35$~g~cm$^{-2}$.  Note that the linear wave at left and the non-linear wave at right are raised by the same perturbing moon, Prometheus; the difference is that the 7:6 resonance is first-order (because $7-6=1$) and thus stronger, while the 15:13 resonance is second-order (because $15-13=2$) and thus weaker. 
\label{SampleSDW}}
\end{center}
\end{figure}

The appearance of the background surface density $\sigma_0$ in \Eqn{}~\ref{WavenumEqn} allows spiral waves to be used diagnostically to measure that parameter at many locations in Saturn's rings.  For waves that conform to the linear theory, the surface density can be taken as constant over the extent of the wave.  Since all other variables in \Eqn{}~\ref{WavenumEqn} have known constant values, we can differentiate with respect to $r$ to obtain a constant value for $\ud k / \ud r$ \citep[see][and references therein]{soirings}.  This tells us that linear waves should produce a slanted feature with a constant slope if we plot $k$ against $r$ (see \Fig{}~\ref{SampleSDW}a, which also introduces the \textit{wavelet transform plots} used throughout this work).  However, most of the prominent waves that appear in Saturn's rings are significantly affected by inter-particle effects such as self-gravity and collisions, which cause them to deviate from linear theory; instead, the signature produced on a wavelet transform plot by these ``non-linear'' waves is concave-up, appearing to hang below the straight slope predicted by linear theory (\Fig{}~\ref{SampleSDW}b).  

\begin{figure}[!t]
\begin{center}
\includegraphics[height=7cm]{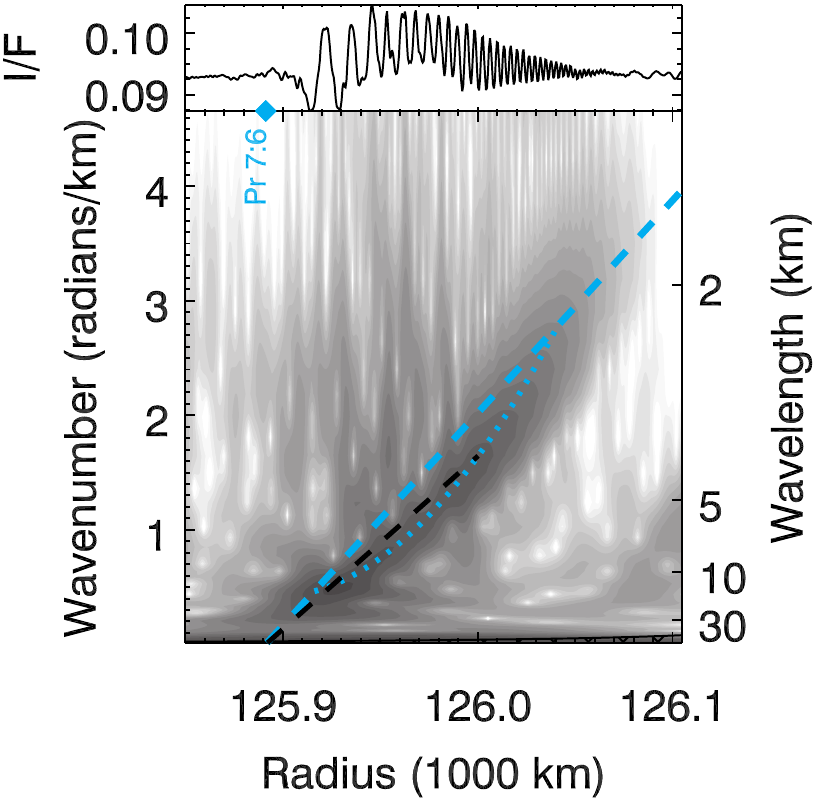}
\caption{The same as \Fig{}~\ref{SampleSDW}b, but with a black dashed line joining the point $(r,k)$ = $(\rres,0)$ to the point $(r,k)$ = $(126~000~\mathrm{km},1.645~\mathrm{km}^{-1})$, which lies along the dotted line.  Plugging these numbers into \Eqn{}~\ref{WavenumEqn} yields $\sigma_0 = 43$~g~cm$^{-2}$, higher than the 35~g~cm$^{-2}$ reflected by the cyan dashed line. 
\label{SampleSDW2}}
\end{center}
\end{figure}

When $\sigma_0$ cannot be assumed to be constant across the wave, then an effective value for $\sigma_0$ at each point $r$ within the wave can be estimated by drawing a line between the wavenumber $k(r)$ and the wave's ``origin'' at $r=\rres$ and $k=0$ (the black dashed line in \Fig{}~\ref{SampleSDW2}).  The concave-up shape of the non-linear wave's trace means that this calculation yields higher values of $\sigma_0$ in the middle parts of the wave.  This likely reflects a real process by which non-linear waves concentrate ring mass within themselves, but it is important to recognize that it does not reflect a true ``background'' surface density, which would be the ring's value of $\sigma_0$ if the wave were not there.  Most \Voyit{}-era measurements of $\sigma_0$ from spiral waves are analogous to the black dashed line in \Fig{}~\ref{SampleSDW2}, and thus are systematically higher than the true background surface density of the ring. 

\begin{figure}[!t]
\begin{center}
\includegraphics[height=3.8cm]{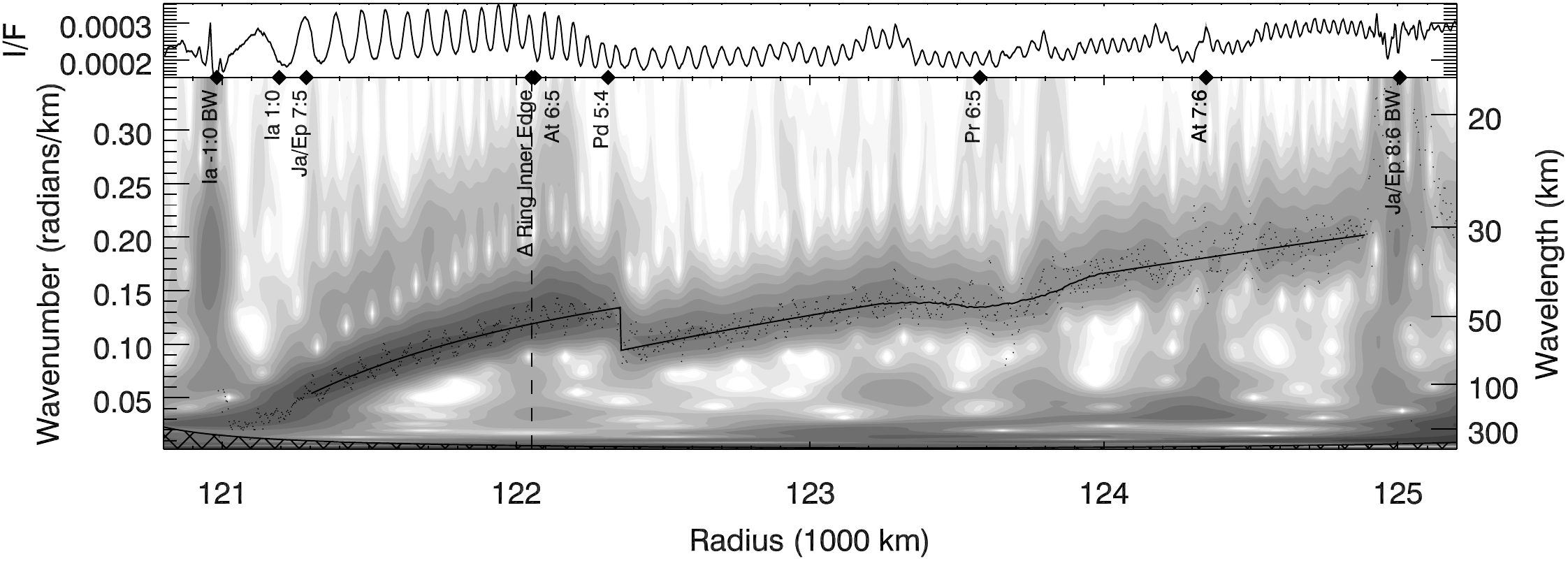}
\hspace{0.5cm}
\includegraphics[height=3.8cm]{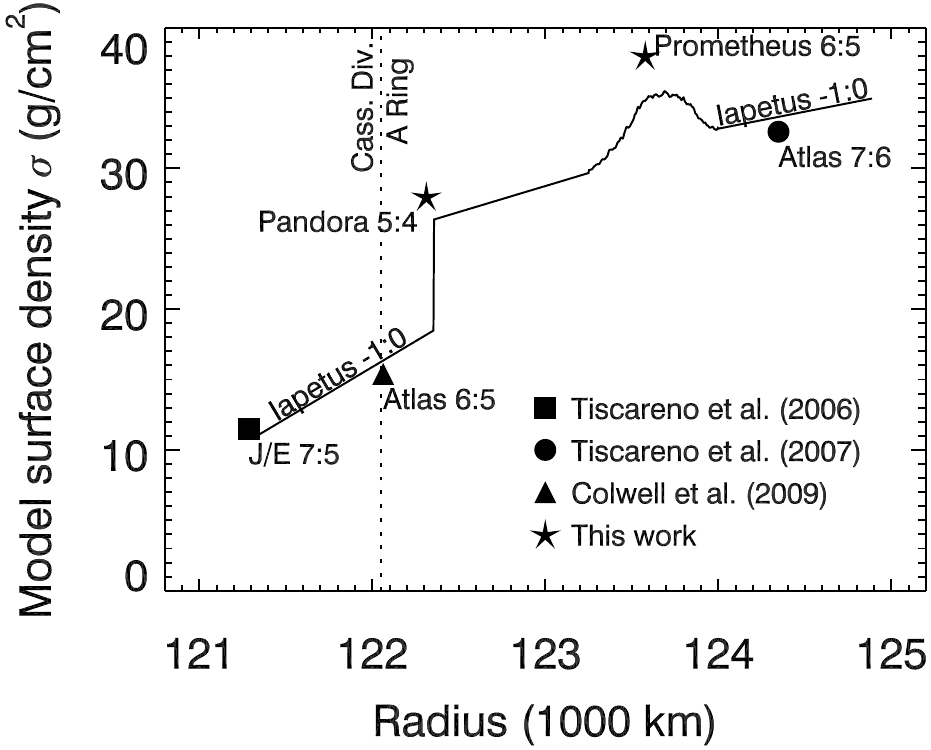}
\caption{Wavelet transform plot (\textit{left}) and derived background surface density profile (\textit{right}) from the Iapetus~-1:0 SBW.  Plots from \citet{Iapetuswave13}. 
\label{IapetusWavePlots}}
\end{center}
\end{figure}

Sometimes a wave extends across a region in which the ring's background surface density naturally varies.  A good example of this is the Iapetus~-1:0 SBW \citep{Iapetuswave13}.  The long wavelengths of this $m=1$ wave (see above) mean that it extends over several thousand~km before it damps away (\Fig{}~\ref{IapetusWavePlots}a), such that plugging the observed series of slopes $k(r)/(r-\rres)$ into \Eqn{}~\ref{WavenumEqn} yields a complex profile of true background surface densities $\sigma_0(r)$, which agrees at several points with $\sigma_0$ values derived from measurements of smaller SDWs (\Fig{}~\ref{IapetusWavePlots}b).  A key factor that makes this possible is that even strong SBWs do not concentrate mass (since their perturbations are perpendicular to the ring plane, rather than within it), so non-linear effects are negligible. 

\subsection{Resonance torques \label{Restorques}}

The amplitude of a SDW or SBW is proportional to the rate at which angular momentum is transferred into the ring (that is, to the torque exerted) by the perturbing moon via the resonance.  The derivation of these resonance torques (or, equivalently, resonance strengths) first appeared in several papers by \citeauthor{GT82} and was summarized by \citet{GT82}.  The locations and strengths of particular resonances occurring within Saturn's main rings were tabulated by \citet{LC82}, whose plot of torque vs. location in the A~ring served as an initial ``atlas'' of the ring's main resonant features.  A primary goal of the present work is to create a series of plots updating the work of \citet{LC82}, which we will present in Section~\ref{Results}. 

As derived by \citet{GT78b}, the torque exerted by a ($m$+$k$):($m$-1) resonance can be expressed as
\begin{equation}
\label{TorqueEqn}
T_{m,k} = - \frac{m \pi^2 \sigma_0}{\mathscr{D}}  \left( \rres \frac{\ud \phi_{m,k}}{\ud a} + \frac{2 n \phi_{m,k}}{|n-n_{m,k}|} \right)^2 , 
\end{equation}
\noindent where $n_{m,k}$ is the resonance's pattern speed and $\phi_{m,k}$ is the term in the Disturbing Function that corresponds to the resonance, both of which are further described below.  


\begin{table}[!t]
\caption{Saturn gravity constants used in this work. \label{saturn_constants}}
\begin{scriptsize}
\begin{center}
\begin{tabular} { c c }
\hline
$GM_\mathrm{Sat}$ & 37931207.7~km$^{3}$~s$^{-2}$ \\
$R_\mathrm{Sat}$ & 60330~km \\
$J_2$ & $16290.71 \times 10^{-6}$ \\
$J_4$ & $-935.83 \times 10^{-6}$ \\
$J_6$ & $86.14 \times 10^{-6}$ \\
$J_8$ & $10 \times 10^{-6}$ \\
\hline
\multicolumn{2}{l}{Source: \citet{Jake06}}
\end{tabular}
\end{center}
\end{scriptsize}
\end{table}

We calculate the mean motion $n$ and the radial frequency $\kappa$, both at the resonance location $\rres$, using the first four spherical harmonic components of Saturn's gravity field (Table~\ref{saturn_constants}.  The mean motion is given by \citep[see \Eqn{s}~6.241, 6.218, and~4.32 of][hereafter MD99]{MD99}
\begin{equation}
n^2 = \frac{GM_\mathrm{Sat}}{a^3} \left[ 1 + \frac{3}{2} J_2 \left( \frac{R_\mathrm{Sat}}{a} \right)^2 - \frac{15}{8} J_4 \left( \frac{R_\mathrm{Sat}}{a} \right)^4 + \frac{35}{16} J_6 \left( \frac{R_\mathrm{Sat}}{a} \right)^6 - \frac{315}{128} J_8 \left( \frac{R_\mathrm{Sat}}{a} \right)^8 \right] , 
\end{equation}
\noindent the radial frequency is given by (see \Eqn{s}~6.242, 6.218, and 4.32 of MD99)
\begin{equation}
\kappa^2 = \frac{GM_\mathrm{Sat}}{a^3} \left[ 1 - \frac{3}{2} J_2 \left( \frac{R_\mathrm{Sat}}{a} \right)^2 + \frac{45}{8} J_4 \left( \frac{R_\mathrm{Sat}}{a} \right)^4 - \frac{175}{16} J_6 \left( \frac{R_\mathrm{Sat}}{a} \right)^6 + \frac{2205}{128} J_8 \left( \frac{R_\mathrm{Sat}}{a} \right)^8 \right] , 
\end{equation}
\noindent and the vertical frequency is given by (see \Eqn{s}~6.243, 6.218, and 4.32 of MD99)
\begin{equation}
\nu^2 = \frac{GM_\mathrm{Sat}}{a^3} \left[ 1 + \frac{9}{2} J_2 \left( \frac{R_\mathrm{Sat}}{a} \right)^2 - \frac{75}{8} J_4 \left( \frac{R_\mathrm{Sat}}{a} \right)^4 + \frac{245}{16} J_6 \left( \frac{R_\mathrm{Sat}}{a} \right)^6 - \frac{2835}{128} J_8 \left( \frac{R_\mathrm{Sat}}{a} \right)^8 \right] . 
\end{equation}

Expressions of $\phi_{m,k}$ for relevant resonances are given below.  We calculate them from the expansion of the Disturbing Function given in Appendix~B of MD99, using an alphanumeric identifier for each term as used in that work.  The calculations require the use of the Laplace coefficients $b_{1/2}^m(\beta)$ for $\beta \equiv \rres/a_\mathrm{s}$, and their derivatives, which we obtain via a numerical algorithm using the method of \citet[see \Eqn{s}~47 through~80 in Chapter~XV of that work]{BC61}.  The following expressions for $\phi_{m,k}$ will use the operators $\mathrm{D} \equiv (\ud/\ud \beta)$ and $\mathrm{D}^n \equiv (\ud^n/\ud^n \beta)$.  

\subsubsection{Inner Lindblad resonances with $m>1$ \label{ILR}}

The pattern speed for a ($k$+1)th-order inner Lindblad resonance (ILR), which drives a spiral density wave (SDW) that propagates outward, is given by
\begin{equation}
\label{PatternSpeedILR}
n^\mathrm{L}_{m,k} = n_\mathrm{s} + \frac{k}{m} \kappa_\mathrm{s} . 
\end{equation}
\noindent The resonance condition is 
\begin{equation}
\label{ResonanceConditionILR}
m n^\mathrm{L}_{m,k} - m n = - \kappa ,
\end{equation}
The LHS of this equation is the forcing frequency, which can be expressed as the derivative of a resonance argument $\varphi^\mathrm{L}_{m,k}$:
\begin{equation}
\label{ResonanceArgumentILR}
\dot{\varphi}^\mathrm{L}_{m,k} = (m+k) n_\mathrm{s} - m n - k \dot{\varpi}_\mathrm{s} , 
\end{equation}
\noindent since the apsidal precession frequency $\dot{\varpi} \equiv n - \kappa$. 

The Disturbing Function (DF) term corresponding to a ($k$+1)th-order ILR is obtained by substituting $j=m+k$ into term 4D$k$.($k$+1) of the MD99 expansion (for example, the term for a first-order resonance is obtained by substituting $j=m$ into argument 4D0.1).  The DF terms are as follows:
\begin{equation}
\label{DW0}
\phi_{m,0}^L = - \frac{GM_\mathrm{s}}{a_\mathrm{s}} \cdot b_{1/2}^m . 
\end{equation}
\begin{equation}
\label{DW1}
\phi_{m,1}^L = - \frac{GM_\mathrm{s}}{a_\mathrm{s}} \cdot e_\mathrm{s} \cdot \left[ \frac{1}{2} + m + \frac{1}{2} \beta \mathrm{D} \right] b_{1/2}^m . 
\end{equation}
\begin{equation}
\label{DW2}
\phi_{m,2}^L = - \frac{GM_\mathrm{s}}{a_\mathrm{s}} \cdot \frac{e_\mathrm{s}^2}{8} \cdot \Big[ (4m^2+9m+4) + (4m+5) \beta \mathrm{D} + \beta^2 \mathrm{D}^2 \Big] b_{1/2}^m . 
\end{equation}
\begin{equation}
\label{DW3}
\phi_{m,3}^L = - \frac{GM_\mathrm{s}}{a_\mathrm{s}} \cdot \frac{e_\mathrm{s}^3}{48} \cdot \Big[ (8m^3+42m^2+65m+27) + (12m^2+51m+51) \beta \mathrm{D} \\ \nonumber
\end{equation}
\vspace{-0.5cm}
\begin{equation}
\hspace{3cm} + (6m+15) \beta^2 \mathrm{D}^2 + \beta^3 \mathrm{D}^3 \Big] b_{1/2}^m . 
\end{equation}
\begin{equation}
\label{DW4}
\hspace{-3cm} \phi_{m,4}^L = - \frac{GM_\mathrm{s}}{a_\mathrm{s}} \cdot \frac{e_\mathrm{s}^4}{384} \cdot \Big[ (16m^4+152m^2+499m^2+646m+256) \\ \nonumber
\end{equation}
\vspace{-0.5cm}
\begin{equation}
\hspace{3.8cm} + (32m^3+264m^2+692m+568) \beta \mathrm{D} \\ \nonumber
\end{equation}
\vspace{-0.5cm}
\begin{equation}
\hspace{2.6cm} + (24m^2+150m+228) \beta^2 \mathrm{D}^2 \\ \nonumber
\end{equation}
\vspace{-0.5cm}
\begin{equation}
\hspace{2.9cm} + (8m+28) \beta^3 \mathrm{D}^3 + \beta^4 \mathrm{D}^4 \Big] b_{1/2}^m . 
\end{equation}
\noindent In addition to being proportional to the mass of the perturbing satellite and inversely proportional to its distance, these terms are also proportional to the perturbing satellite's eccentricity $e_\mathrm{S}$ to the $k$th power (meaning that the first-order term is not proportional to $e_\mathrm{s}$ at all).  

The expressions given by \Eqn{s}~\ref{DW0} through~\ref{DW2} are identical to those written down by \citet{GT82}, while \Eqn{s}~\ref{DW3} and~\ref{DW4} extend the coverage to fifth-order SDWs.  A more generalized treatment is desirable and will be the subject of future work. 

To facilitate plugging \Eqn{s}~\ref{DW0} through~\ref{DW4} into \Eqn{}~\ref{TorqueEqn}, we also write the derivatives:
\begin{equation}
\frac{\ud \phi_{m,0}^L}{\ud a} = - \frac{GM_\mathrm{s}}{a_\mathrm{s} \rres} \cdot \beta \mathrm{D} b_{1/2}^m . 
\end{equation}
\begin{equation}
\frac{\ud \phi_{m,1}^L}{\ud a} = - \frac{GM_\mathrm{s}}{a_\mathrm{s} \rres} \cdot e_\mathrm{s} \cdot \left[ (1+m) \beta \mathrm{D} + \frac{1}{2} \beta^2 \mathrm{D}^2 \right] b_{1/2}^m . 
\end{equation}
\begin{equation}
\frac{\ud \phi_{m,2}^L}{\ud a} = - \frac{GM_\mathrm{s}}{a_\mathrm{s} \rres} \cdot \frac{e_\mathrm{s}^2}{8} \cdot \Big[ (4m^2+13m+9) \beta \mathrm{D} + (4m+7) \beta^2 \mathrm{D}^2 + \beta^3 \mathrm{D}^3 \Big] b_{1/2}^m . 
\end{equation}
\begin{equation}
\frac{\ud \phi_{m,3}^L}{\ud a} = - \frac{GM_\mathrm{s}}{a_\mathrm{s} \rres} \cdot \frac{e_\mathrm{s}^3}{48} \cdot \Big[ (8m^3+54m^2+116m+78) \beta \mathrm{D} + (12m^2+63m+81) \beta^2 \mathrm{D}^2 \\ \nonumber
\end{equation}
\vspace{-0.5cm}
\begin{equation}
\hspace{3cm} + (6m+18) \beta^3 \mathrm{D}^3 + \beta^4 \mathrm{D}^4 \Big] b_{1/2}^m . 
\end{equation}
\begin{equation}
\hspace{-3cm} \frac{\ud \phi_{m,4}^L}{\ud a} = - \frac{GM_\mathrm{s}}{a_\mathrm{s} \rres} \cdot \frac{e_\mathrm{s}^4}{384} \cdot \Big[ (16m^4+184m^3+763m^2+1338m+824) \beta \mathrm{D} \\ \nonumber
\end{equation}
\vspace{-0.5cm}
\begin{equation}
\hspace{4cm} + (32m^3+312m^2+992m+1024) \beta^2 \mathrm{D}^2 \\ \nonumber
\end{equation}
\vspace{-0.5cm}
\begin{equation}
\hspace{2.3cm} + (24m^2+174m+312) \beta^3 \mathrm{D}^3 \\ \nonumber
\end{equation}
\vspace{-0.5cm}
\begin{equation}
\hspace{2.6cm} + (8m+32) \beta^4 \mathrm{D}^4 + \beta^5 \mathrm{D}^5 \Big] b_{1/2}^m . 
\end{equation}

\subsubsection{Inner Lindblad resonances with $m=1$, \label{ILRm1}}
The pattern speed for a $m=1$ ILR is the same as in \Eqn{}~\ref{PatternSpeedILR}, $n_{1,0}^\mathrm{L} = n_\mathrm{s}$, but now \Eqn{}~\ref{ResonanceConditionILR} reduces to $\varpi = n_\mathrm{s}$, which means that the apsidal precession frequency of the ring particle resonates with the mean motion of the perturbing moon, with the mean motion of the ring particle no longer taking part.  The DF term now includes the indirect term (4E0.1 in MD99) and becomes
\begin{equation}
\phi_{m,0}^L = - \frac{GM_\mathrm{s}}{a_\mathrm{s}} \cdot \Big[ b_{1/2}^m - \beta \Big] . 
\end{equation}
\noindent with
\begin{equation}
\frac{\ud \phi_{m,0}^L}{\ud a} = - \frac{GM_\mathrm{s}}{a_\mathrm{s} \rres} \cdot \Big[ \beta \mathrm{D} b_{1/2}^m - \beta \Big] . 
\end{equation}

\subsubsection{Outer Lindblad resonances \label{OLR}}
An outer Lindblad resonance (OLR) occurs when the resonance location is outward of the perturbing moon, which in Saturn's rings occurs only for resonances with Pan, though it would also occur for resonances with Daphnis if any were observed.  In this case the SDW propagates inward.  The resonance condition changes sign, which we denote by retaining \Eqn{}~\ref{ResonanceConditionILR} but defining $m < 0$.  The remaining calculations are then analogous to those in Section~\ref{ILR}. 

\subsubsection{Vertical resonances \label{VR}}
The pattern speed for a ($k$+1)th-order inner vertical resonance (IVR), which drives a spiral bending wave (SBW) that propagates inward, is given by
\begin{equation}
\label{PatternSpeedIVR}
n^\mathrm{V}_{m,k} = n_\mathrm{s} + \frac{k-1}{m} \kappa_\mathrm{s} + \frac{1}{m} \nu_\mathrm{s} . 
\end{equation}
\noindent The resonance condition is 
\begin{equation}
\label{ResonanceConditionIVR}
m n^\mathrm{V}_{m,k} - m n = - \nu ,
\end{equation}
and the derivative of the resonance argument $\varphi^\mathrm{V}_{m,k}$ is
\begin{equation}
\label{ResonanceArgumentIVR}
\dot{\varphi}^\mathrm{V}_{m,k} = (m+k) n_\mathrm{s} - m n - (k-1) \dot{\varpi}_\mathrm{s} - \dot{\Omega}_\mathrm{s} , 
\end{equation}
\noindent since the nodal precession frequency $\dot{\Omega} \equiv n - \nu$. 

Expressions for the torque of a vertical resonance are less clearly stated in the literature than analogous expressions for Lindblad resonances.  It is desirable to write down a clearer and more generalized theory of vertical resonances and SBWs, but we leave this for future work.  For the purposes of the resonance atlases in Section~\ref{Results}, we will locate each vertical resonance on the plot by making use of torque for the associated Lindblad resonance.  

Outer vertical resonances (OVRs) do not appear in Saturn's rings, as Pan is the only wave-driving moon with ring material exterior to it, and Pan has no measurable orbital inclination \citep{Jake08}.  Therefore, we will set aside OVRs for the present time. 

\subsection{Orbital elements of perturbing moons}
All parameters of moon orbits that are needed for calculating properties of resonances within Saturn's main rings are given in Table~\ref{orbelems}.  These values were calculated by analyzing the best-fit orbits made available on the website\fn{\texttt{ftp://naif.jpl.nasa.gov/pub/naif/generic\_kernels}} of NASA's Navigation and Ancillary Information Facility (NAIF).  The kernels that were used in this work are listed in Table~\ref{kertable}. 

The parameters listed in Table~\ref{orbelems} are consistent with the orbital elements given by \citet{Jake08} and on the JPL~website\fn{\texttt{https://ssd.jpl.nasa.gov/?sat\_elem}}.  However, the values given here include the precession frequencies $\dot{\Omega}$ and $\dot{\varpi}$ explicitly, rather than relying on an assumption (which does not always hold) that the precession frequencies can be straightforwardly calculated from the mean motion.  Furthermore, the values given here include variation in each parameter over the life of the \Cassit{}~mission.  A more detailed analysis of the changing orbits of Saturn's moons will be given in a forthcoming paper, but the values given here will suffice for calculating resonances to the precision needed for analyzying \Cassit{} images. 

We here explain three items in Table~\ref{orbelems} that might catch the reader's eye:  The semimajor axis of Hyperion varies widely due to the 4:3 mean-motion resonance with Titan.  The inclination of Iapetus \textit{as seen from the rings} does not precess (that is, $\dot{\Omega} = 0$) because the Laplace plane at Iapetus' distance is dragged away from Saturn's equatorial plane due to the gravity of the Sun \citep{TTN09}.  The orbit of Pan is assumed to be circular and within Saturn's equatorial plane, which does not much matter because Pan's mass is too small for any second-order resonances to be observable in any case. 

\begin{table}[!t]
\begin{center}
\caption{Orbital elements for Saturn's moons, derived from primary analysis of NAIF trajectory kernels.  Ranges (given after the $\pm$ sign) are not uncertainties but actual variation of the parameter over the life of the \Cassit{} mission. \label{orbelems}}
\begin{scriptsize}
\begin{tabular}{l c r @{~$\pm$~} l r @{~$\pm$~} l r @{~$\pm$~} l r @{~$\pm$~} l c c }
\hline
\hline
Moon & NAIF ID & \multicolumn{2}{c}{$n$, $^\circ$~d$^{-1}$} & \multicolumn{2}{c}{$a$, km} & \multicolumn{2}{c}{$e$} & \multicolumn{2}{c}{I, $^\circ$} & $\dot{\Omega}$, $^\circ$~d$^{-1}$ & $\dot{\varpi}$, $^\circ$~d$^{-1}$ \\
\hline
Mimas & 601 &  381.985 &  0.009 &   185537 &      3 & 0.0196 & 0.0002 &  1.565 & 0.003 &   -0.999 &    1.001 \\
Enceladus & 602 &  262.732 &  0.005 &   238035 &      3 & 0.0047 & 0.0001 &  0.009 & 0.005 &   -0.416 &    0.338 \\
Tethys & 603 &  190.698 &  0.005 &   294672 &      5 & 0.0002 & 0.0001 &  1.091 & 0.002 &   -0.198 &    0.199 \\
Dione & 604 &  131.535 &  0.005 &   377410 &     10 & 0.0022 & 0.0002 &  0.03 & 0.01 &   -0.081 &    0.084 \\
Rhea & 605 &   79.690 &  0.007 &   527070 &     30 & 0.0009 & 0.0002 &  0.36 & 0.01 &   -0.025 &   -0.001 \\
Titan & 606 &   22.577 &  0.001 &  1222160 &     50 & 0.0287 & 0.0003 &  0.38 & 0.03 &   -0.000 &    0.001 \\
Hyperion & 607 &   16.92 &  0.09 &  1480800 &   5500 & 0.10 & 0.02 &  1.1 & 0.2 &   -0.004 &   -0.058 \\
Iapetus & 608 &    4.5379 &  0.0001 &  3560560 &     50 & 0.030 & 0.005 & 15.6 & 0.4 & 0 &    0.001 \\
Janus (a) & 610 &  518.345 &  0.005 &   151441 &      1 & 0.0068 & 0.0001 &  0.164 & 0.001 &   -2.046 &    2.054 \\
Janus (b) & 610 &  518.238 &  0.005 &   151462 &      1 & 0.0068 & 0.0001 &  0.164 & 0.001 &   -2.045 &    2.053 \\
Epimetheus (a) & 611 &  518.099 &  0.005 &   151489 &      1 & 0.0097 & 0.0001 &  0.352 & 0.001 &   -2.044 &    2.052 \\
Epimetheus (b) & 611 &  518.485 &  0.005 &   151414 &      1 & 0.0097 & 0.0001 &  0.352 & 0.001 &   -2.047 &    2.055 \\
Atlas & 615 &  598.31 &  0.01 &   137666 &      2 & 0.0012 & 0.0001 &  0.003 & 0.001 &   -2.868 &    2.881 \\
Prometheus & 616 &  587.285 &  0.006 &   139378 &      1 & 0.00223 & 0.00001 &  0.007 & 0.001 &   -2.745 &    2.758 \\
Pandora & 617 &  572.789 &  0.012 &   141713 &      2 & 0.00419 & 0.00004 &  0.050 & 0.001 &   -2.588 &    2.600 \\
Pan & 618 &  626.032 & 0 &   133584 &      0 & 0 & 0 & 0 & 0 & 0 & 0 \\
\hline
\multicolumn{12}{l}{For Janus and Epimetheus, configuration (a) is from Feb~2006 to Feb~2010 and again from Feb~2014 to Feb~2018, while}\\
\multicolumn{12}{l}{configuration (b) is from Feb~2002 to Feb~2006 and again from Feb~2010 to Feb~2014.}
\end{tabular}
\end{scriptsize}
\end{center}
\end{table}

\begin{table}[!t]
\begin{center}
\caption{NAIF kernels used in orbital analysis for this work. \label{kertable}}
\begin{scriptsize}
\begin{tabular}{l c l }
\hline
\hline
Moon & NAIF ID & Kernels \\
\hline
Mimas & 601 & de430, sat365 \\
Enceladus & 602 & de430, sat365 \\
Tethys & 603 & de430, sat365 \\
Dione & 604 & de430, sat365 \\
Rhea & 605 & de421, sat359L \\
Titan & 606 & de421, sat359L \\
Hyperion & 607 & de421, sat359L \\
Iapetus & 608 & de421, sat359L \\
Janus & 610 & de421, sat357 \\
Epimetheus & 611 & de421, sat357 \\
Atlas & 615 & de430, sat363 \\
Prometheus & 616 & de430, sat363 \\
Pandora & 617 & de430, sat363 \\
Pan & 618 & de430, sat363-rocks-merge \\
\hline
\multicolumn{3}{l}{Also used for all moons were cas00166 and naif0011.}
\end{tabular}
\end{scriptsize}
\end{center}
\end{table}

\begin{table}[!t]
\begin{center}
\caption{Masses for Saturn's moons. \label{masses}}
\begin{scriptsize}
\begin{tabular}{l c l l }
\hline
\hline
Moon & NAIF ID & Mass, kg & Reference \\
\hline
Mimas & 601 & $3.749 \times 10^{19}$ & \citet{Jake06} \\
Enceladus & 602 & $1.080 \times 10^{20}$ & \citet{Jake06} \\
Tethys & 603 & $6.174 \times 10^{20}$ & \citet{Jake06} \\
Dione & 604 & $1.0955 \times 10^{21} $ & \citet{Jake06} \\
Rhea & 605 & $2.3065 \times 10^{21}$ & \citet{Jake06} \\
Titan & 606 & $1.3452 \times 10^{23}$ & \citet{Jake06} \\
Hyperion & 607 & $5.62 \times 10^{18}$ & \citet{Thomas10} \\
Iapetus & 608 & $1.8056 \times 10^{21}$ & \citet{Jake06} \\
Janus & 610 & $1.8975 \times 10^{18}$ & \citet{Thomas10} \\
Epimetheus & 611 & $5.266 \times 10^{17}$ & \citet{Thomas10} \\
Atlas & 615 & $6.6 \times 10^{15}$ & \citet{Thomas10} \\
Prometheus & 616 & $1.60 \times 10^{17}$ & \citet{Thomas10} \\
Pandora & 617 & $1.37 \times 10^{17}$ & \citet{Thomas10} \\
Pan & 618 & $5.0 \times 10^{15}$ & \citet{Thomas10} \\
\hline
\end{tabular}
\end{scriptsize}
\end{center}
\end{table}

\subsection{Masses of perturbing moons}
The masses of relevant moons of Saturn are given in Table~\ref{masses}.  They are determined either through dynamical analysis of each moon's gravitational effects on the orbits of neighboring moons \citep{Jake06} or by determining the moon's size and shape and assuming a density \citep{Thomas10}. 

\section{Observations \label{Observations}}

Operational details of the \Cassit{}~ISS camera, including the calibration protocol, were described by \citet{PorcoSSR04}

We will refer to individual \Cassit{}~ISS images by their 10-digit identifier (e.g., N1560310964), where the initial ``N'' signifies the Narrow Angle Camera (NAC).  We will refer to groups of images that were obtained together for a common scientific purpose by their 3-digit orbit number and their 9-character observation identifier.  For example, N1560310964 is part of an observation called 046/RDHRESSCN (a coded representation of ``high-resolution radial scan,'' those these are not standardized), signifying that the images were obtained during \Cassit{}'s 46th orbit\fn{Actually its 47th, since the \Cassit{} mission began with Orbits~A, B, and~C, followed by Orbit~003.  This was the result of a last-minute redesign of the beginning of the orbital tour in order to enable increased observations of Titan in advance of the descent of the Huygens lander.} around Saturn and were intended to be a high-resolution radial scan\fn{Here we speak of a ``radial scan'' of the rings, consisting of a series of images whose fields of view range from the inner edge to the outer edge of the rings.  More often in this work we will speak of a ``radial scan'' of an image (see Section~\ref{RadialScans}), which is a one-dimensional array of brightness values as a function of radial location, obtained by averaging over the points within the image that share each radial location.  It is unfortunate that the same phrase is used for these two different concepts.} of the rings. 

\begin{figure}[!t]
\begin{center}
\includegraphics[width=16cm]{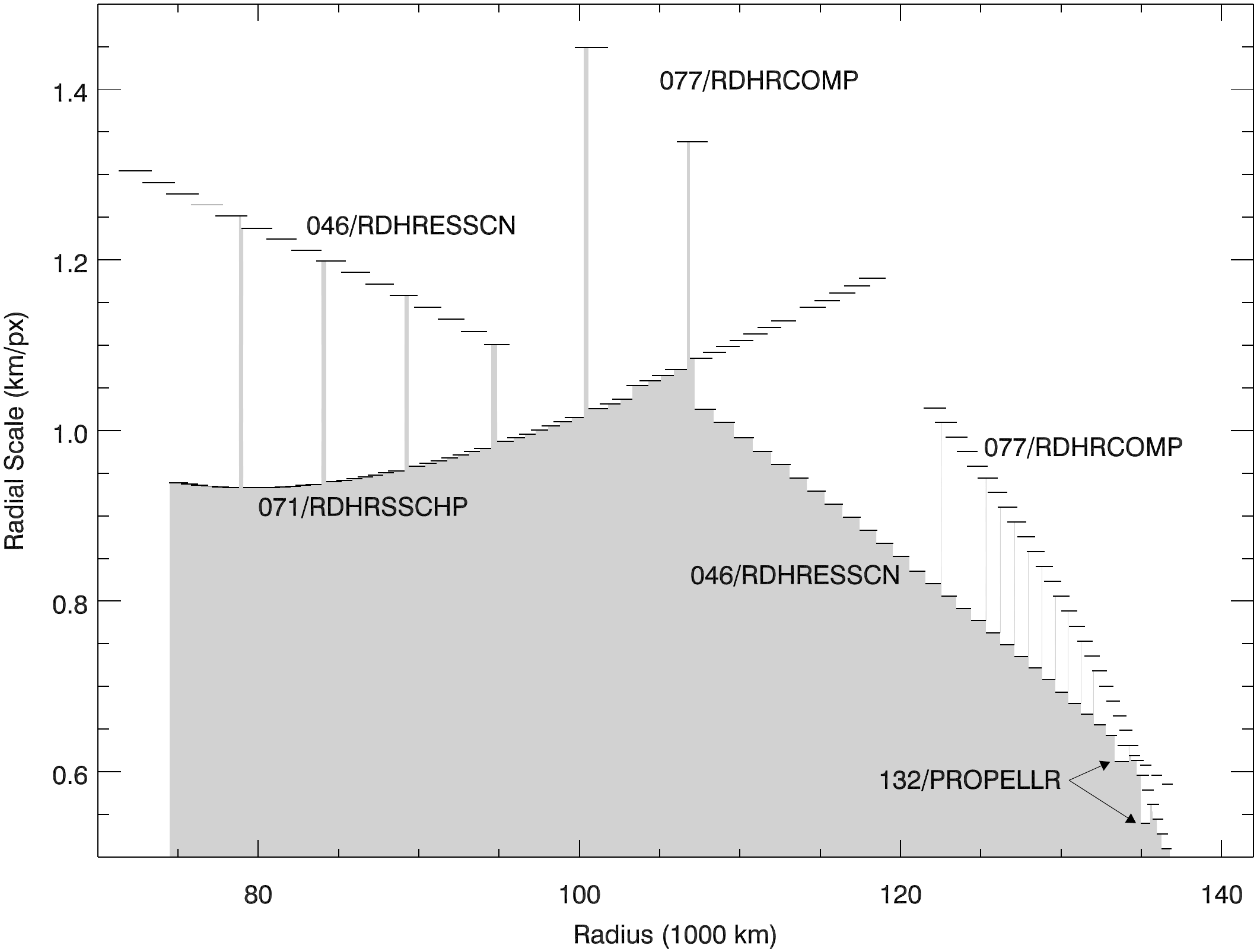}
\caption{Schematic representation of the resolution (pixel scale) at which each location within the main rings is covered by images in the current data set.  Every horizontal dash represents one image in our data set.  For each location ($x$-axis), the gray region rises upward to the resolution of the best image capturing that location.  In the C~ring, the best images are generally from 071/RDHRSSCHP (obtained 9~June~2008); however, some of these images do not overlap with each other, so there are regions for which 046/RDHRESSCN is needed to fill gaps.  Similarly, in the A~ring, the best images are generally from 046/RDHRESSCN (obtained 12~June~2007), but 077/RDHRCOMP is needed to fill gaps between images.  In the outermost A~ring, 077/RDHRCOMP (obtained 21~July~2008) and 132/PROPELLR (obtained 3~June~2010) have superior resolution.  Each individual image represented here is shown in Appendix~B as \Fig{s}~A7 
through~A140
 . 
\label{RingsresRadialScales}}
\end{center}
\end{figure}

This work focuses on the highest-quality images of the main rings obtained by Cassini after Saturn Orbit Insertion (SOI) and before the Ring Grazing Orbits (RGOs) and Grand Finale (GF).  The RGO and GF images were obtained after this work was completed but before it was published.  The SOI images, though they have much better nominal resolutions (many around 0.3~km~pixel$^{-1}$) are much noisier due to the short exposures required to prevent smear while the spacecraft was moving very rapidly in a lateral direction across the rings.  The result is that the images in the current data set are comparable if not better in discernible radial structure (for example, in the wavelet scans as shown in Appendix~B) compared to the SOI images. 

The observations used are 046/RDHRESSCN, 071/RDHRSSCHP, 077/RDHRCOMP, and \linebreak 132/PROPELLR (\Fig{}~\ref{RingsresRadialScales}).  The image numbers, nominal radial scale, and phase angles of the images in the current data set are given in the figure captions of Appendix~B (see also Table~\ref{ObsParams}).  

\begin{table}[!t]
\begin{center}
\caption{Observation parameters for data sets used in this work. \label{ObsParams}}
\begin{scriptsize}
\begin{tabular}{l c c c }
\hline
\hline
 & Best Radial Pixel & & \\
Observation ID & Scale, km pixel$^{-1}$ & Phase Angle & Hour Angle \\
\hline
046/RDHRESSCN & 0.59 & 36.1$^\circ$ -- 48.3$^\circ$ & 251.5$^\circ$ -- 271.0$^\circ$ \\
071/RDHRSSCHP & 0.93 & 77.4$^\circ$ -- 120.6$^\circ$ & 210.5$^\circ$ -- 233.6$^\circ$ \\
077/RDHRCOMP & 0.51 & 81.1$^\circ$ -- 100.1$^\circ$ & 142.6$^\circ$ -- 154.0$^\circ$ \\
132/PROPELLR & 0.54 & 24.4$^\circ$ -- 58.1$^\circ$ & 267.3$^\circ$ -- 306.4$^\circ$ \\
\hline
\end{tabular}
\end{scriptsize}
\end{center}
\end{table}

Because spiral bending waves (SBWs) appear as corrugations in the ring plane that are close to azimuthal in their direction, their visibility depends on the observed location's ring longitude ($\lambda$) relative to the Sun's longitude in the ring plane ($\lambda_\odot$).  If the hour angle $\lambda - \lambda_\odot$ is near 90$^{\circ}$ (dusk) or 270$^{\circ}$ (dawn), then the Sun's light is shining along the corrugations and SBW signatures will be muted.  If $\lambda - \lambda_\odot$ is near 0$^{\circ}$ (noon) or 180$^{\circ}$ (midnight, but obviously not close enough for the observed location to be within Saturn's shadow), then the Sun's light is shining across the corrugations and SBW signatures will be more prominent.  It happens that both 071/RDHRSSCHP and 077/RDHRCOMP observe locations near the boundary of Saturn's shadow, while 046/RDHRESSCN observes locations near dawn (Table~\ref{ObsParams}), so SBWs are expected to be more prominent in the 071 and 077 observations and less prominent in the 046 observations. 

\section{Methods \label{Methods}}

\begin{figure}[!t]
\floatbox[{\capbeside\thisfloatsetup{capbesideposition={right,top},capbesidewidth=7cm}}]{figure}[\FBwidth]
{\caption{A portion of N1560310219, from the ``high-resolution radial scan'' 046/RDHRESSCN, shown here as an example (analysis is shown in \Fig{}~\ref{RadialScanDemo2}).  The nominal radial scale for this image is   0.7~km~pixel$^{-1}$.  As indicated by the annotations, the ring radius (i.e., the distance from Saturn's center) increases from bottom to top.  The most prominent features in this image are the outward-propagating Prometheus~12:11 spiral density wave, and the inward-propagating Mimas~5:3 spiral bending wave.  The rest of the image is pervaded by more subtle structure that resembles the grooves on a vinyl record.}
\label{RadialScanDemo1}}
{\includegraphics[width=8cm]{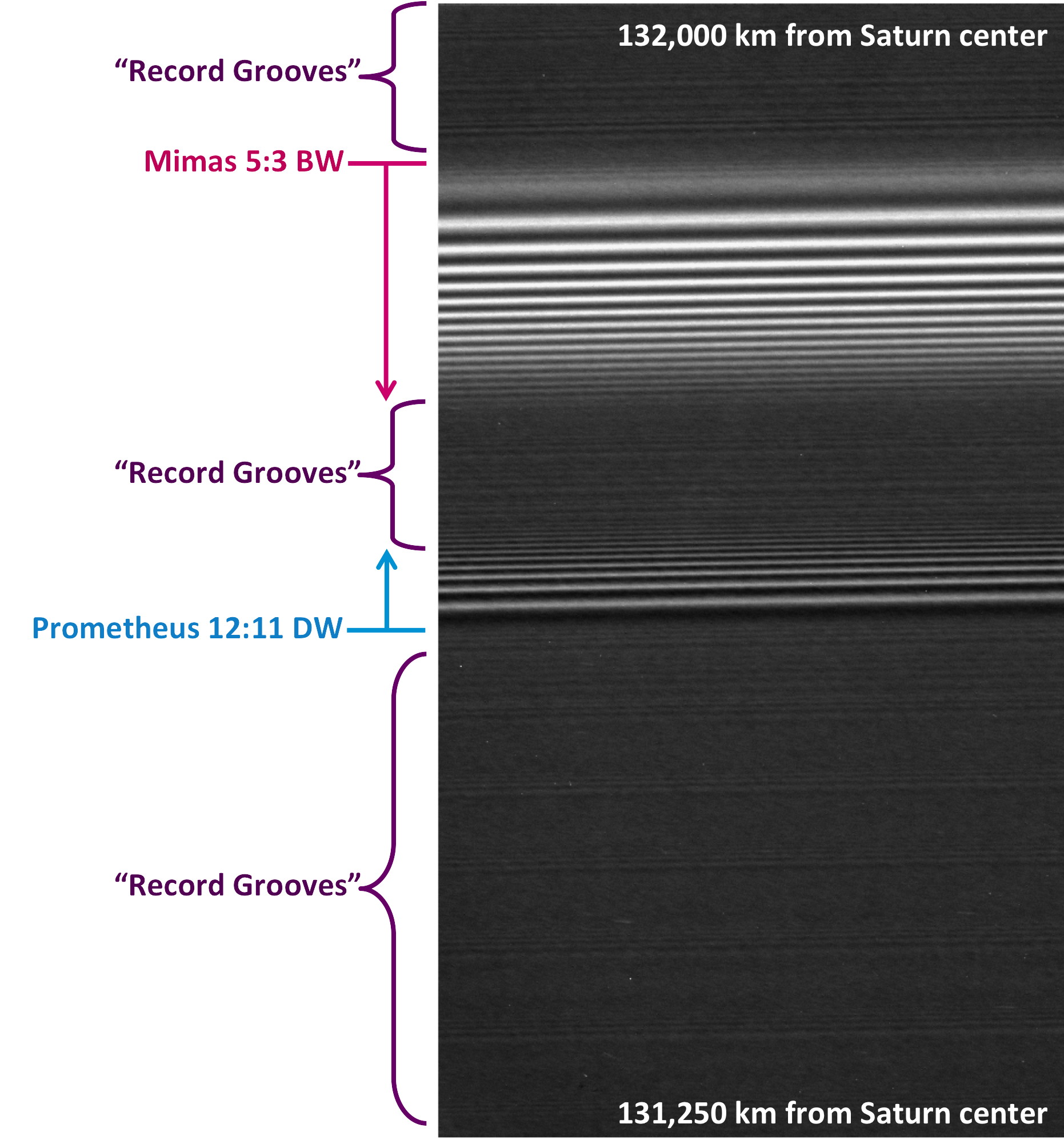}}
\end{figure}

\begin{figure}[!t]
\begin{center}
\vspace{1cm}
\includegraphics[width=16cm,keepaspectratio=true]{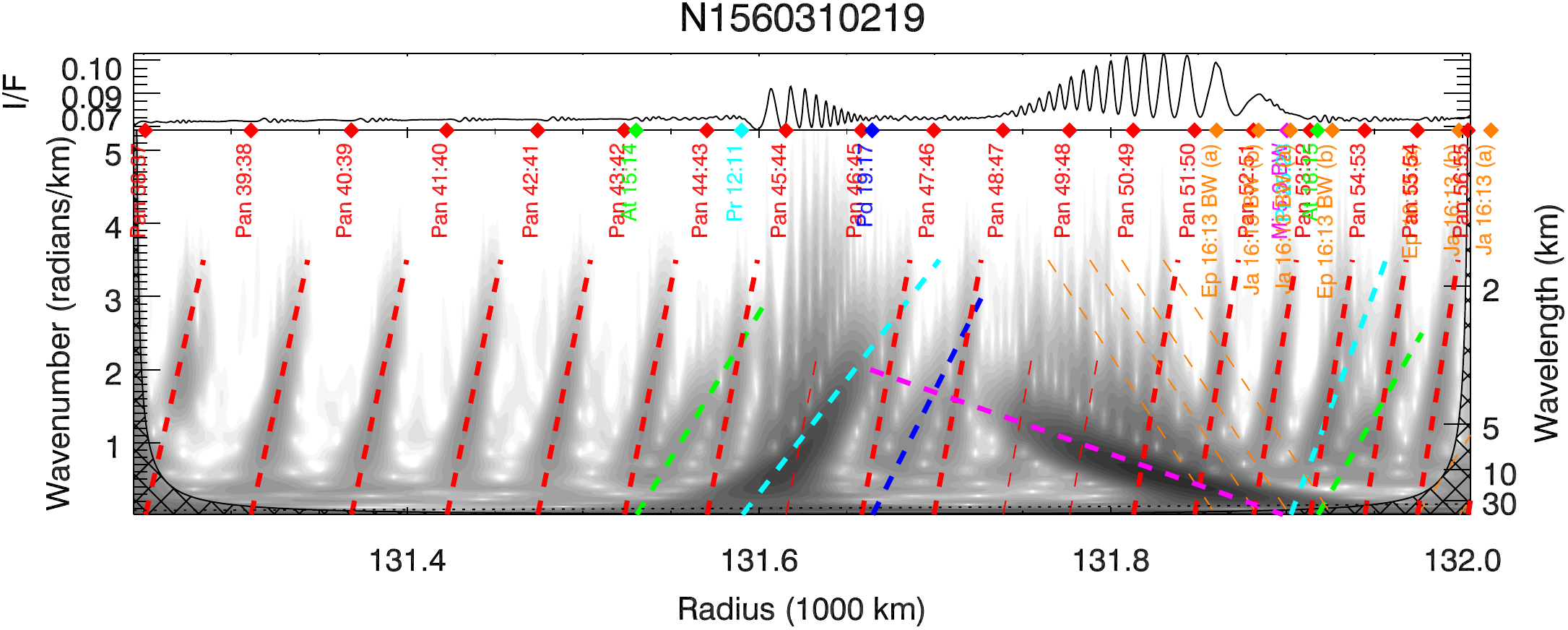}
\caption{Radial scan and wavelet plot for image N1560310219, continuing the example begun in \Fig{}~\ref{RadialScanDemo1}.  As described in \Fig{}~\ref{SampleSDW}, the radial scan extends across the top, and the grayscale shading in the main part of the plot indicates power in the wavelet transform (Section~\ref{RadialScans}).  The signature of the Prometheus~12:11~SDW hangs downward from the associated cyan dashed line with a concave-up geometry (Section~\ref{Dispersion}), while the signature of the Mimas~5:3~SBW hews more closely to the associated magenta dashed line, because SBWs do not display significant non-linearity.  The ``record grooves'' mentioned in \Fig{}~\ref{RadialScanDemo1} are here revealed to be weaker SDWs, mostly resonance with Pan (red) but some resonances with Atlas (green) and second order resonances with Prometheus (cyan) and Pandora (dark blue).  Wave models shown in bold are judged to appear in the data, while those shown as thin lines (e.g., Pan~49:48, the Janus/Epimetheus 16:13 bending waves) are not perceptible because they are overwritten by other, strong structures.  All wave models in this figure use a background surface density $\sigma_0 = 32$~g~cm$^{-2}$. 
\label{RadialScanDemo2}}
\end{center}
\end{figure}

\subsection{Radial scans and wavelet plots \label{RadialScans}}

After navigating each image with appropriate SPICE kernels for the spacecraft's trajectory as well as for the positions of relevant celestial bodies (Table~\ref{kertable}), a grid in ring coordinates (radius and longitude within Saturn's equatorial plane) can be laid across the image.  For each radial location, interpolated values across the longitudes represented in the image are averaged, and these are compiled in order to obtain a one-dimensional ``radial scan'' of the image (e.g., the curve across the top of \Fig{}~\ref{RadialScanDemo2} is a radial scan of the image shown in \Fig{}~\ref{RadialScanDemo1}).  

As described in Section~\ref{Dispersion} and \Fig{}~\ref{SampleSDW}, we take the continuous wavelet transform of each radial scan \citep{TC98,Addison02,soirings}, a technique that optimally indicates the power at each spatial wavenumber $k$ at each radial location $r$.  The resulting plots are given in Appendix~B as \Fig{s}~A7 
through~A140
.  The cross-hatched regions in some wavelet plots indicate the ``cone of influence,'' within which the wavelet transform is contaminated by the edges of the image \citep{TC98,soirings}.  

\subsection{Radial navigation \label{RadialNav}}

Available SPICE kernels encoding the orientation of the \Cassit{} spacecraft are reliable only within a fraction of the ISS~NAC field of view (6~mrad).  The navigation for each image must be fine-pointed using the information within the image.  Whenever stars are visible in the image, we use them to verify the pointing \citep{FrenchDPS16}; however, many images lack stars because of the limited transparency of the rings.  We secondly use sharp edges and other ring features that have been tabulated as fiducials \citep{French93,French17}; however, these also are lacking in many images.  

A third way in which an image can be ``anchored'' in an absolute sense is if it contains at least one SBW or OLR-generated SDW, which propagate inward (see Sections~\ref{ILR} through~\ref{VR}).  An image that contains only ILR-generated SDWs (the dominant form of spiral waves), which propagate outward, is difficult to anchor because a radial offset combined with a change in the inferred surface density can yield a fit that looks almost as good as the true fit.  However, such an adjustment will cause an inward-propagating wave to move visibly away from the best fit, leading to a strong verification of the correct pointing. 

Finally, for images for which none of the above-mentioned methods are available, we compare the radial scan to that of other images in our data set.  Most images have radial coverage that overlaps with that of the images obtained before and after it.  Starting with an image that has been anchored by one of the above-mentioned methods, we overplot adjacent radial scans on top of each other and adjust the radial offset of the radial scan (\textit{not} the two-dimensional pointing of the source image) for the non-anchored image until the structures within the scans match.  An image whose pointing is verified in this way is now anchored and can be used to verify the pointing for the next adjacent image.  In a few cases, adjacent images fail to overlap in radial coverage, and we compared them to images obtained at a different time with overlapping radial coverage. 

\subsection{Evaluating the presence of ring features \label{FeatureEval}}

A quasi-periodic feature in a radial scan will create a linear or quasi-linear signature in the wavelet plot.  A true periodic feature with a constant wavelength $\lambda$ will produce a horizontal wavelet signature.  A feature with a wavenumber $k = 2 \pi / \lambda$ increasing (decreasing) at a constant rate will create a linear wavelet signature that slants upward (downward): this is expected for a spiral wave that follows the linear theory.  A non-linear spiral wave will create a wavelet signature that curves or is otherwise non-linear (see Section~\ref{Dispersion}).  A moonlet wake, created when a recently passed nearby moon organizes the streamlines of ring particles, also creates a curved wavelet signature \citep{Show86,soirings}.  

Conversely, bare peaks and sharp edges produce power at all frequencies.  Similarly, an aspect of ``peakiness'' in the wavecrest of a quasi-sinusoidal feature will produce power in frequencies above that of the quasi-sinusoidal frequency.  Finally, dappled or speckled wavelet signatures indicate simple noise. 

We scanned every wavelet plot in this data set according to this rubric.  We began by overlaying the wavelet plot with the expected wave signatures (\Eqn{}~\ref{WavenumEqn}) and evaluating whether the predicted signatures were present or not.  At the same time, we noted any quasi-linear wavelet signatures that were not predicted from known resonances.  Our results are discussed in Section~\ref{IndivFeat}. 

\subsection{Evaluating the local surface density \label{SurfDensEval}}

\begin{figure}[!t]
\begin{center}
\includegraphics[width=7.5cm,keepaspectratio=true]{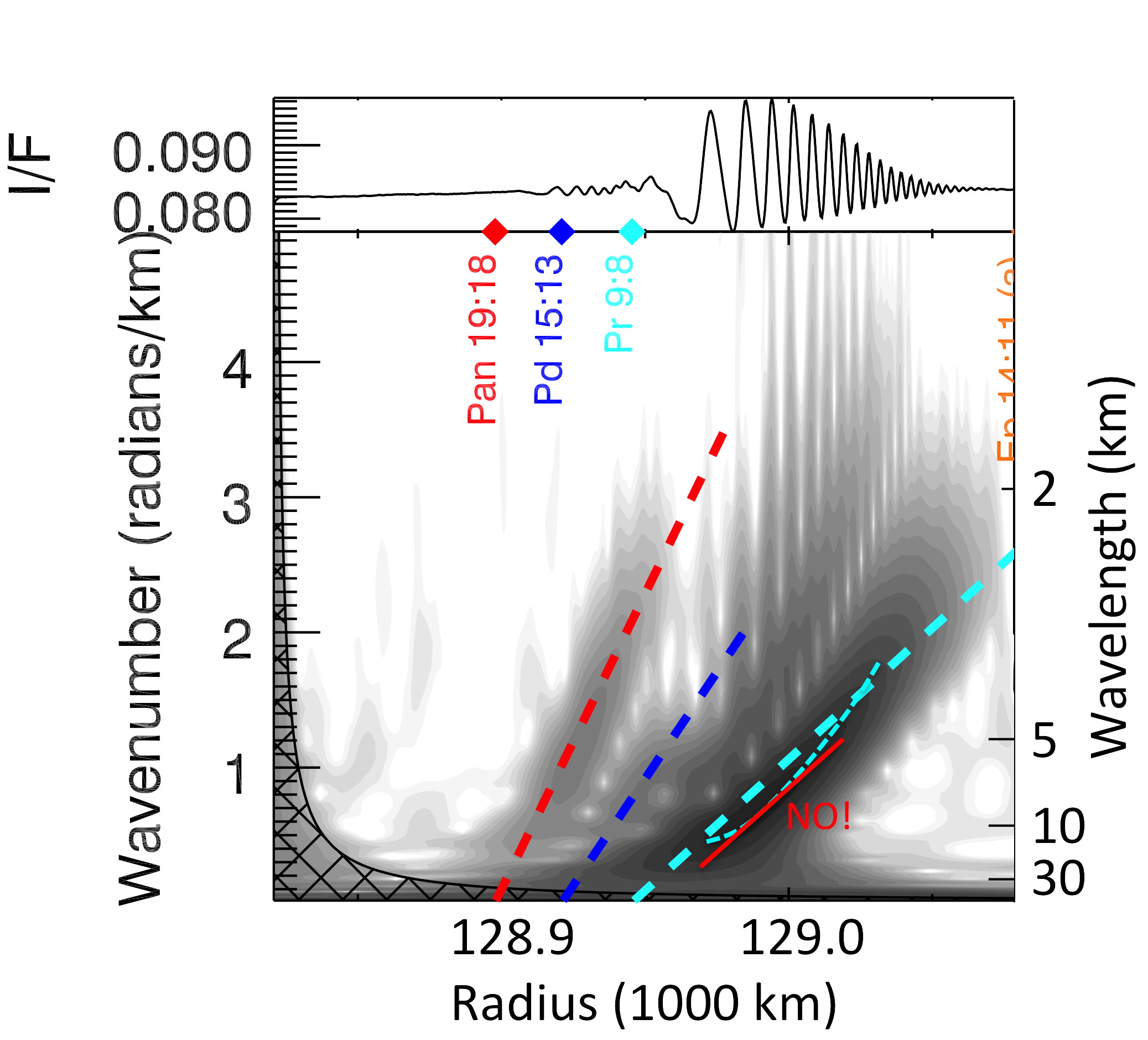}
\hspace{1cm}
\includegraphics[height=6.8cm,keepaspectratio=true]{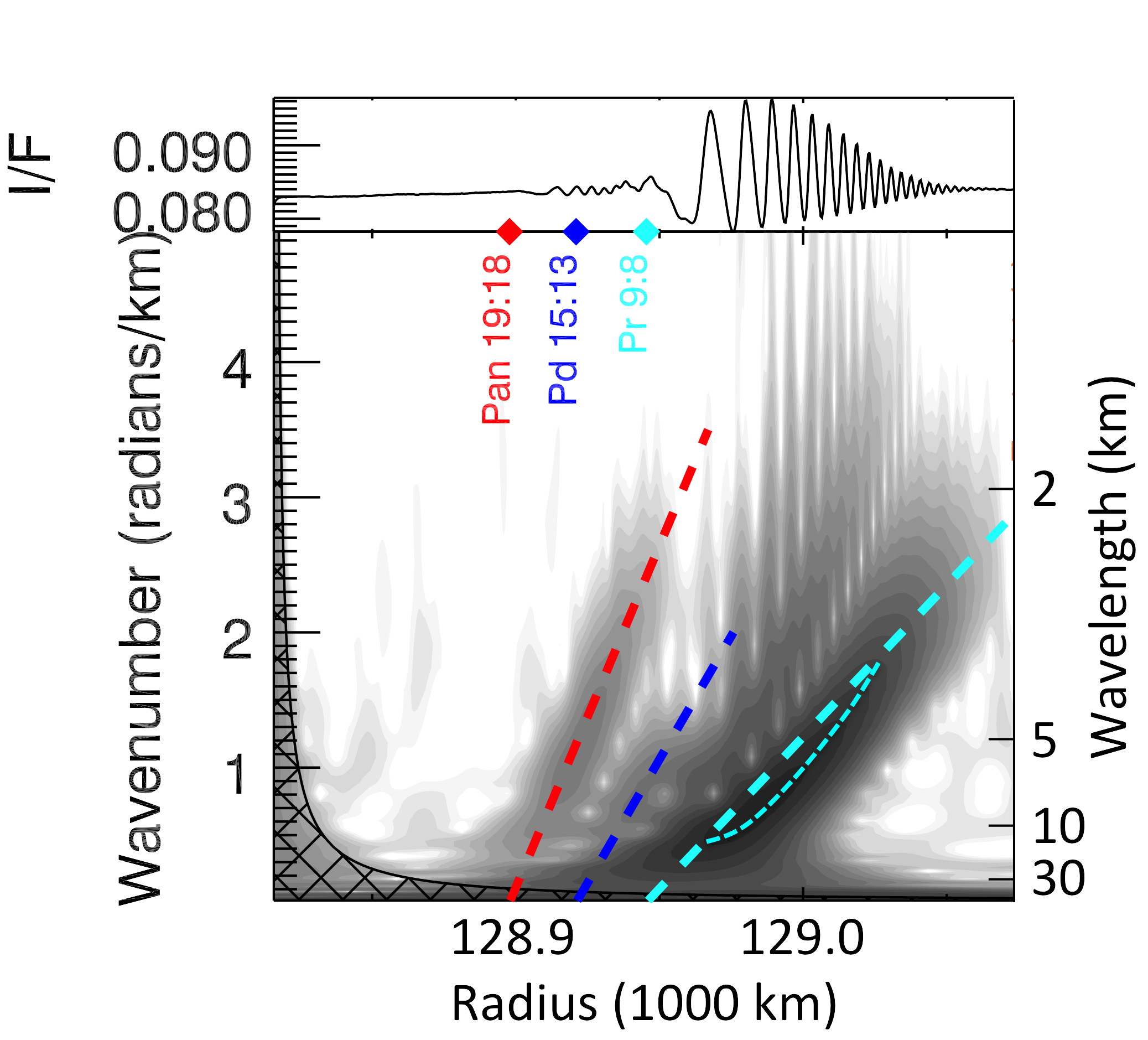}
\caption{A portion of the wavelet plot for image N1560310609, illustrating the wrong and right ways to fit a non-linear SDW.  The cyan dashed line that is thinner and curved has been drawn by hand to trace the actual profile of the wavelet signature of the Prometheus~9:8 SDW, which is concave-up due to non-linear effects (Section~\ref{Dispersion}).  In the left panel, the wave model (the cyan dashed line that is thick and linear) is set at $\sigma_0 = 40.5$~g~cm$^{-2}$ in order to more or less bisect the wavelet signature, which is wrong.  The radial scale has been shifted inward by 15~km in order to ``correctly'' locate the model profiles with their too-shallow slopes, illustrating the importance of proper radial navigation (Section~\ref{RadialNav}).  In the right panel, the wave model is set properly at $\sigma_0 = 34$~g~cm$^{-2}$, and the observed non-linear profile appears to ``hang down'' from the linear model (Section~\ref{Dispersion}).  We suspect that this effect has caused \Voyit{}-era surface density measurements of non-linear SDWs to be consistently higher than those reported in this work (\Fig{}~\ref{SigmaRef3vcass}), and we are confident that this effect causes some values reported by \citet{soirings} for the mid-A~ring to be too high. 
\label{FitExample}}
\end{center}
\end{figure}

Once the radial navigation was established (Section~\ref{RadialNav}), we adjusted the background surface density $\sigma_0$ until the predicted wave model overlay the observed wavelet signatures, according to the principles described in Section~\ref{Dispersion}.  In most cases, it was sufficient to use a single value of $\sigma_0$ for the entire image, as the surface density can be expected to vary smoothly across the ring.  However, in a few cases it was necessary to use multiple $\sigma_0$ values for a single image, as there was evidence for a stronger trend or break in the surface density profile at that location. 

In this work, we did not carry out individual model fits for each spiral wave, as did \citet{soirings}.  Such fits require a clearer wave signature with more wavecrests than does the method used in this work.  Our purpose here is to ascertain the presence or absence of as many features as possible.  The method used here is also able to benefit in many cases from the collective presence of many waves, using them to strengthen confidence in the inferred values. 

The use of all available waves to obtain a surface density value for each locality, along with more careful application of methods to determine the radial navigation and the inferred surface density from non-linear waves (\Fig{}~\ref{FitExample}), yields a surface density profile for the A~ring that has lower values than those previously published, especially in the mid-A~ring.  We argue that this profile (presented in Section~\ref{SurfDens}) corrects previous errors and is much more definitive than any previous work. 

\section{Results \label{Results}}

\subsection{Individual features of interest \label{IndivFeat}}

The results of this work include several classes of waves that (to our knowledge) are here reported for the first time, including waves driven by Enceladus and by Hyperion, second- and third-order SBWs and third-order SDWs driven by Janus and Epimetheus, and several features whose origins are unknown. 

\subsubsection{Waves driven by Hyperion and Enceladus}

\begin{figure}[!t]
\begin{center}
\vspace{0.3cm}
\includegraphics[height=6.8cm,keepaspectratio=true]{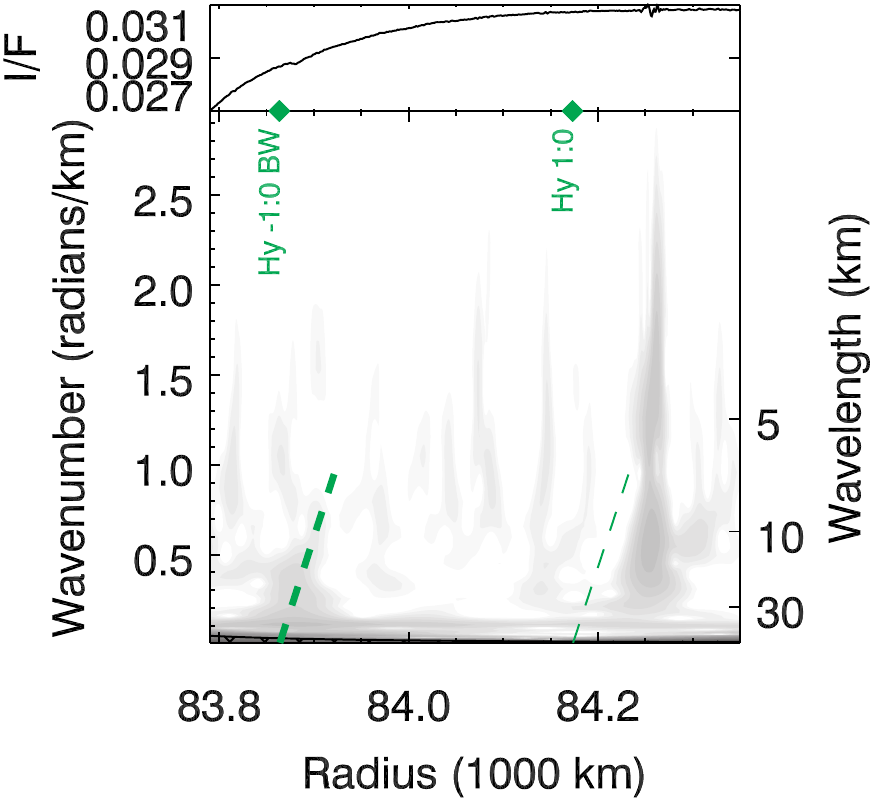}
\caption{A portion of the wavelet plot for image N1560314196, highlighting structure caused by the moon Hyperion.  The full radial scan and wavelet plot for this image is \Fig{}~A46 
in Appendix~B.  The wavelet signature of the Hyperion~$-1$:0 nodal bending wave is present, while that of the Hyperion~1:0 apsidal density wave is not. \label{HypWaves}}
\end{center}
\end{figure}

Like the more massive moons Titan and Iapetus whose orbits lie on either side of it, Hyperion has $m=1$ resonances that fall within the rings.  Here we report that the $-1$:0~nodal bending wave appears in the mid-C~ring, while the 1:0~apsidal density wave does not appear in our data set.  The region in which both resonances lie is shown in \Fig{}~\ref{HypWaves}. 

\begin{figure}[!t]
\begin{center}
\includegraphics[height=6.8cm,keepaspectratio=true]{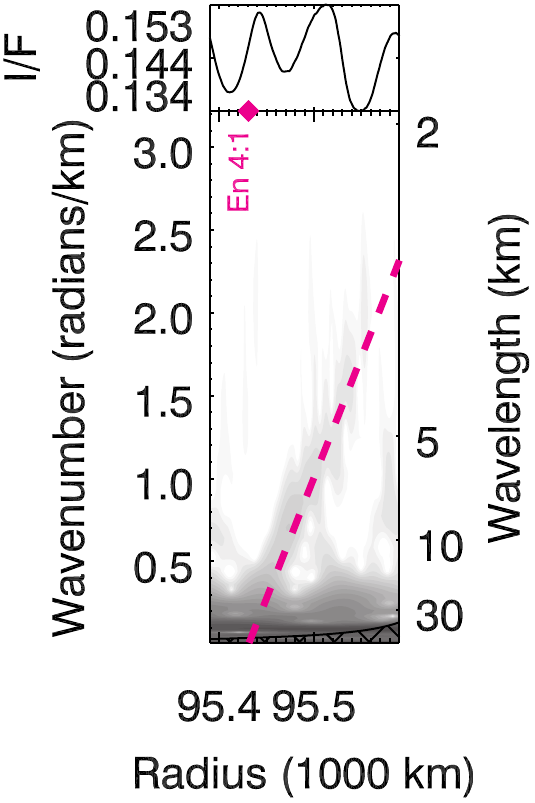}
\includegraphics[height=6.8cm,keepaspectratio=true]{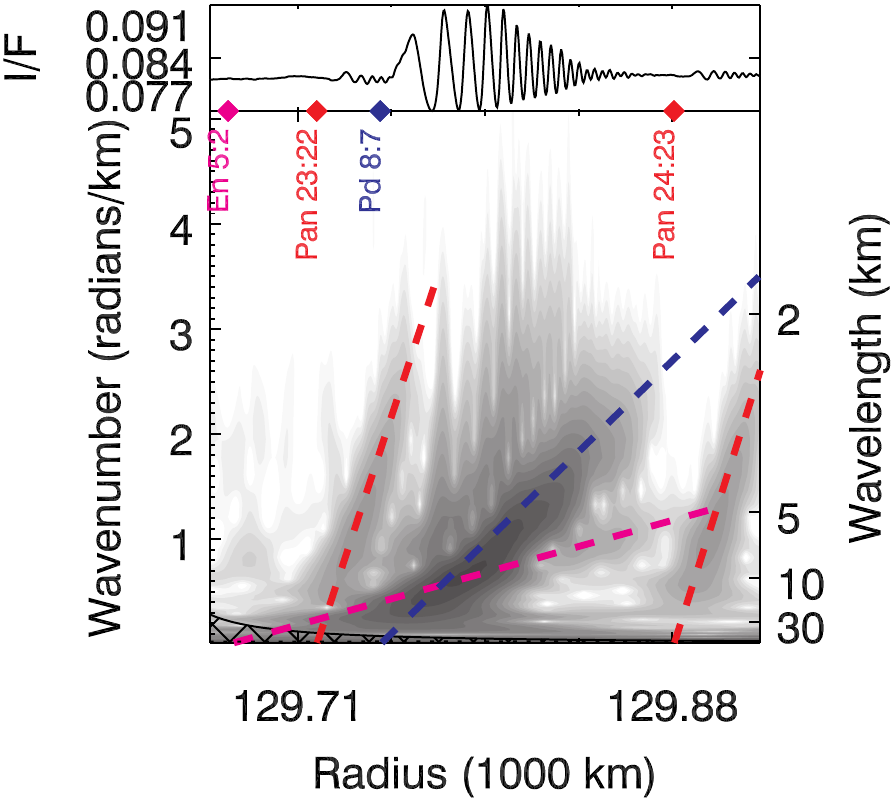}
\caption{Portions of the wavelet plots for images N1560313366 (left) and N1560310460 (right), highlighting structure caused by the moon Enceladus.  The full radial scans and wavelet plots for these images are \Fig{s}~A39 
and~A16 
in Appendix~B, respectively.  The Enceladus~4:1 SDW appears in the inner~B ring.  The upstream portion of the Enceladus~5:2 SDW is difficult to discern due to the presence of the stronger Pandora~8:7 SDW; however, downstream it appears with some separation due to its longer wavelengths. \label{EncWaves}}
\end{center}
\end{figure}

Enceladus is the innermost moon for which no first-order resonances fall within the rings (its 2:1 occurring at 150,474~km).  The only second-order Enceladus resonance to all within the rings is 3:1, which falls within the B~ring at 115,207~km and is not discernible in any single image due to other B~ring structure that overwrites it, though it has been discerned by combining the signals from many VIMS occultations \citep{HN16}.  Two third-order Enceladus waves, however, are clearly discernible in our data set: the 4:1~SDW in the inner B~ring and the 5:2~SDW in the mid-A~ring (\Fig{}~\ref{EncWaves}).  

\subsubsection{Waves driven by Janus and Epimetheus}

\begin{figure}[!t]
\begin{center}
\includegraphics[height=6.8cm,keepaspectratio=true]{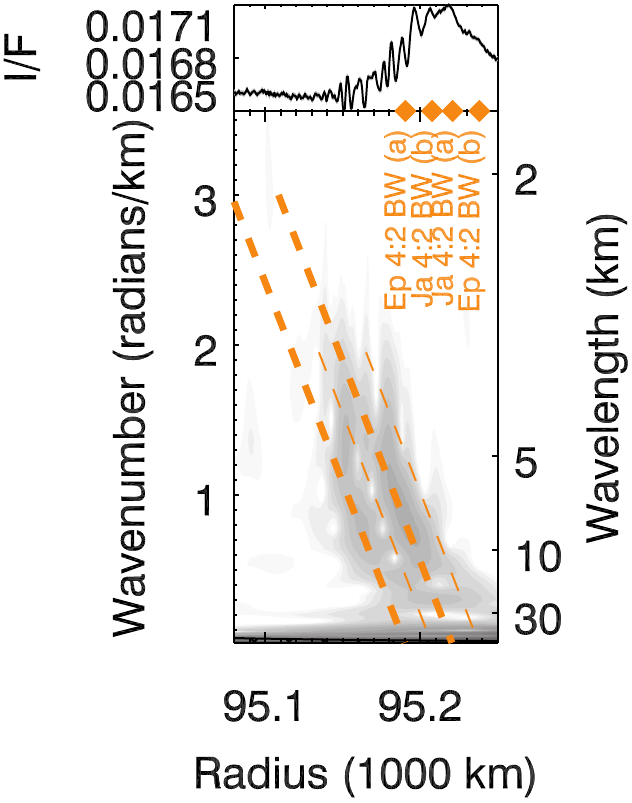}
\includegraphics[height=6.8cm,keepaspectratio=true]{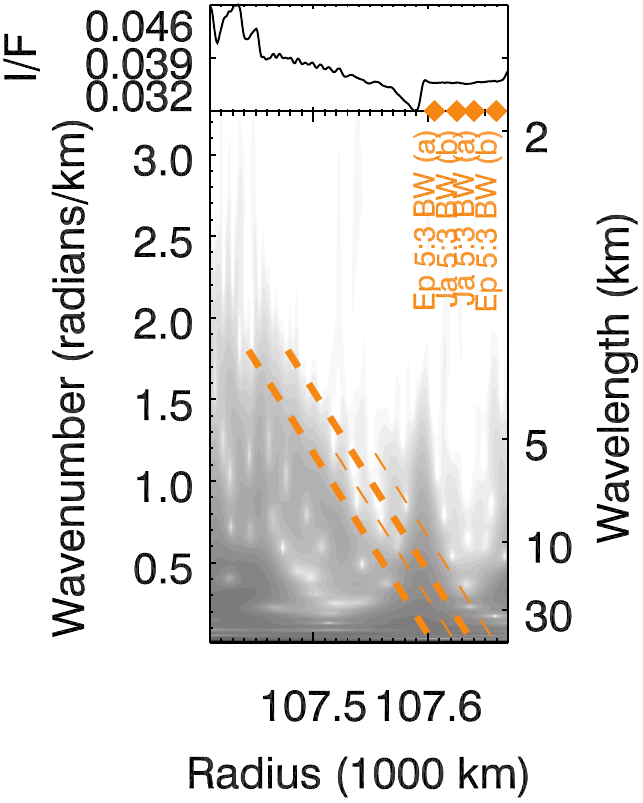}
\vspace{0.5cm}
\includegraphics[height=6.8cm,keepaspectratio=true]{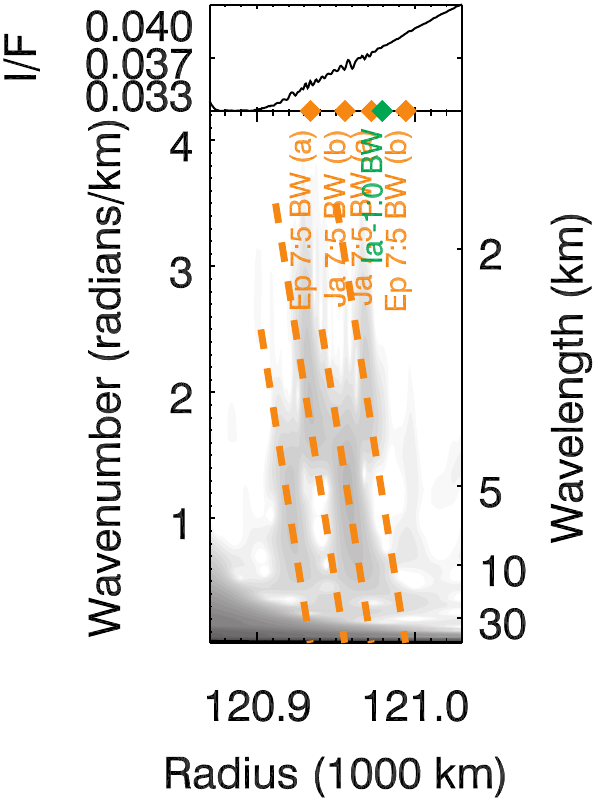}
\includegraphics[height=6.8cm,keepaspectratio=true]{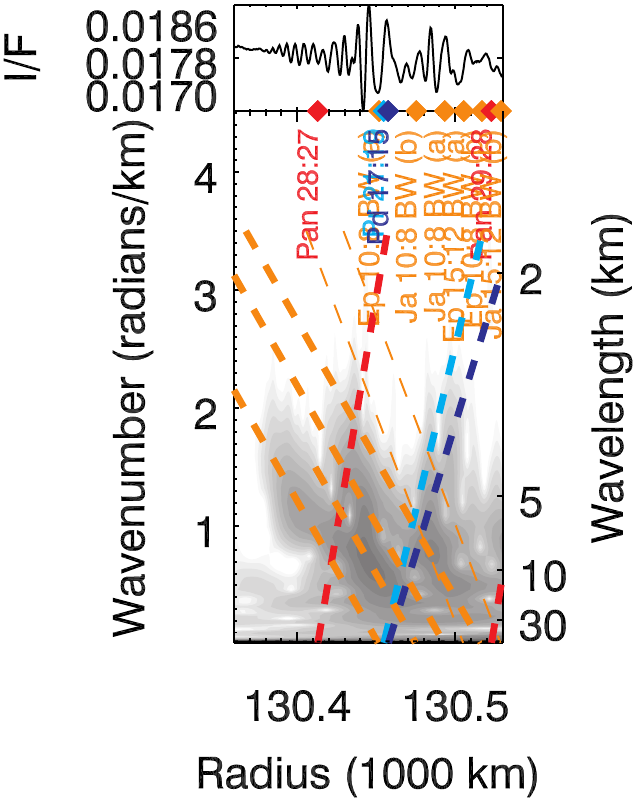}
\includegraphics[height=6.8cm,keepaspectratio=true]{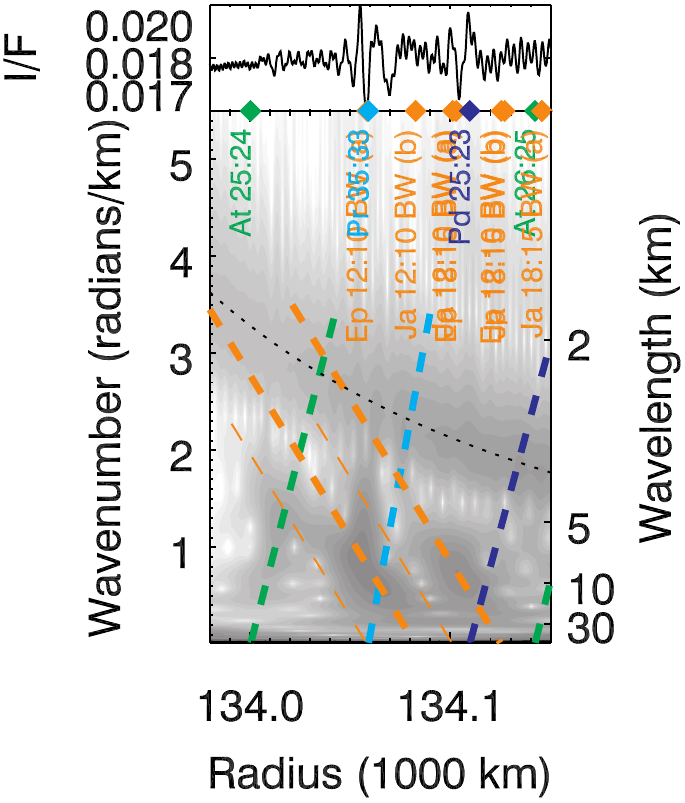}
\includegraphics[height=6.8cm,keepaspectratio=true]{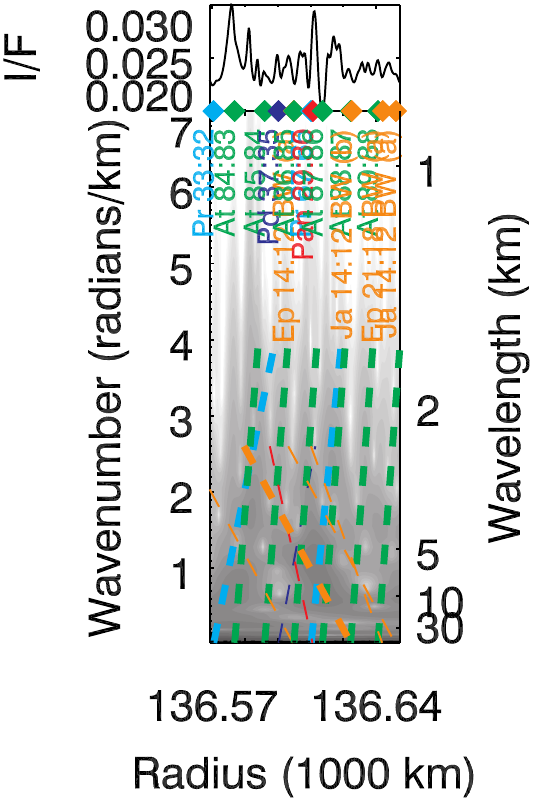}
\caption{Portions of the wavelet plots for images N1591678028 (top-left), N1591679726 (top-middle), N1560311666 (top-right), N1595338974 (bottom-right), N1595337686 (bottom-middle), and N1595336397 (bottom-right), highlighting second-order SBWs caused by the co-orbital moons Janus and Epimetheus.  The full radial scans and wavelet plots for these images are \Fig{s}~A80
, A94
, A26
, A122
, A114
, and A106 
in Appendix~B, respectively.  In addition to the second-order Janus/Epimetheus SBWs shown here, the 8:6, 9:7, and 13:11 are shown in \Fig{}~\ref{UnexA}.  \label{JE2}}
\end{center}
\end{figure}

\begin{figure}[!t]
\begin{center}
\includegraphics[height=6.8cm,keepaspectratio=true]{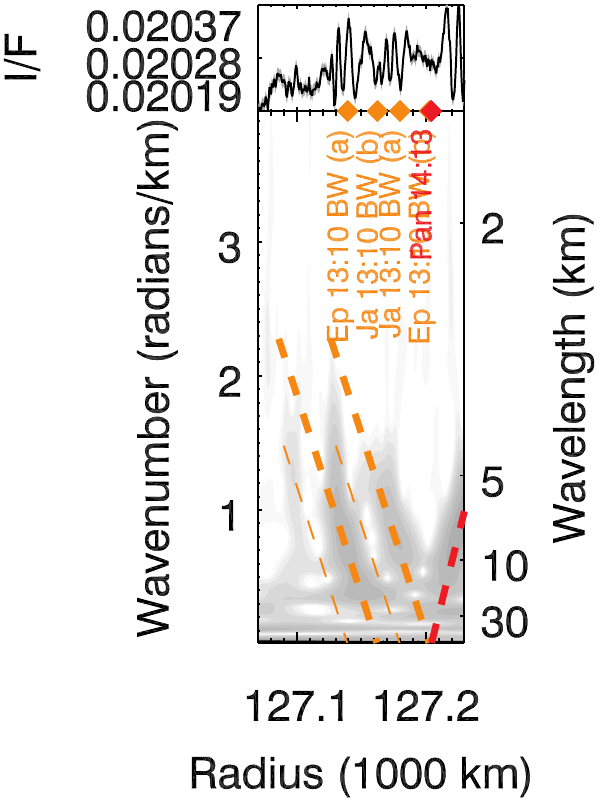}
\vspace{0.5cm}
\includegraphics[height=6.8cm,keepaspectratio=true]{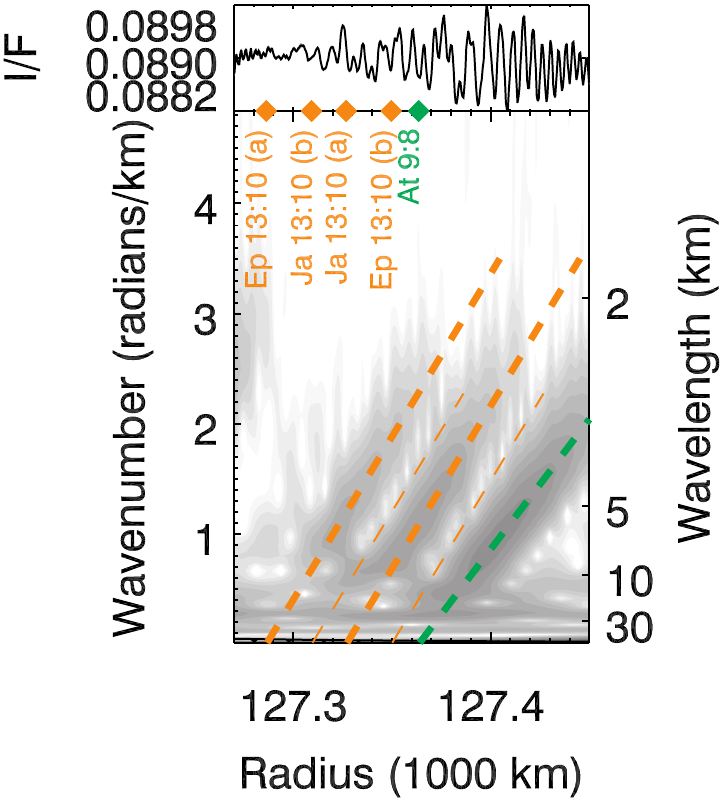}
\includegraphics[height=6.8cm,keepaspectratio=true]{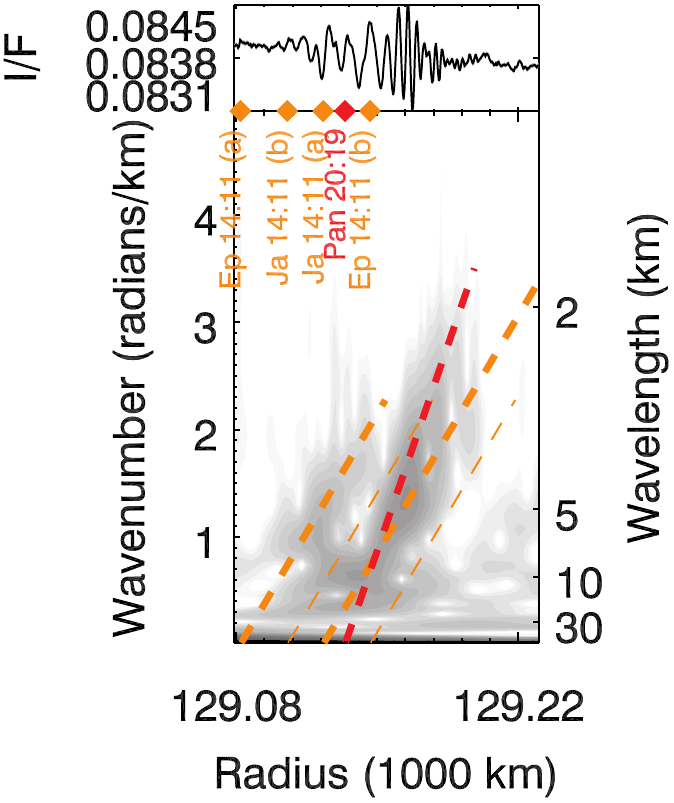}
\includegraphics[height=6.8cm,keepaspectratio=true]{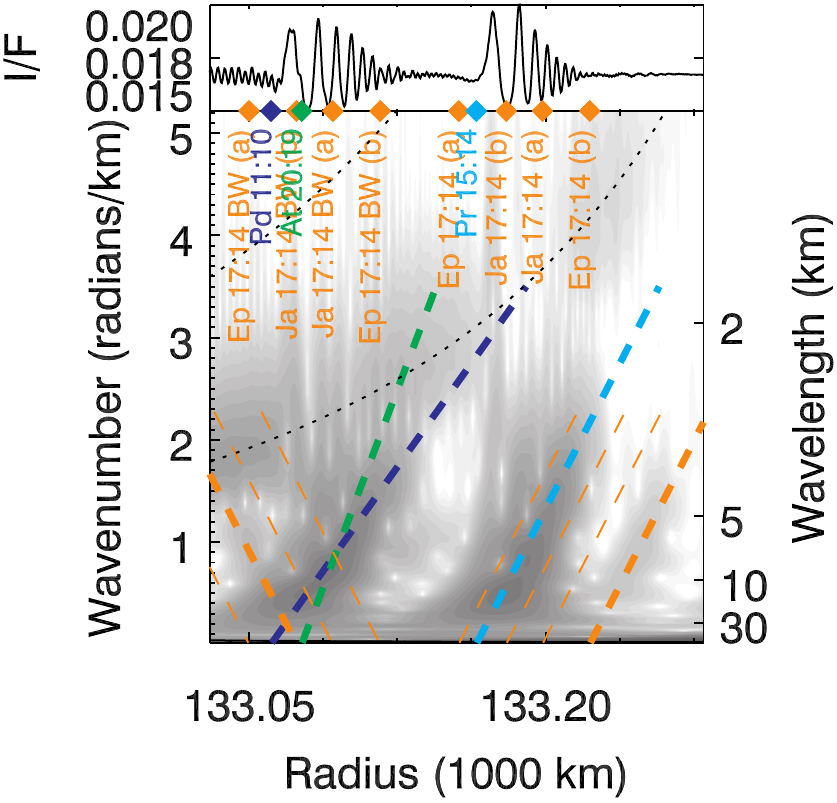}
\caption{Portions of the wavelet plots for images N1595339940 (top-left), N1560310846 (top-right), N1560310609 (bottom-left), and N1595338008 (bottom-right), highlighting third-order SBWs and SDWs caused by the co-orbital moons Janus and Epimetheus.  The full radial scans and wavelet plots for these images are \Fig{s}~A128
, A19
, A17
, and A116 
in Appendix~B, respectively. \label{JE3}}
\end{center}
\end{figure}

Second-order SDWs driven by Janus and Epimetheus were reported by \citet{jemodelshort,soirings}, but not SBWs of either second- or third-order.  A single third-order Janus/Epimetheus SDW, the 17:14, was noted by \citet{soirings}, but here we report many more.  

Because Janus and Epimetheus swap orbits every 4.00 Earth years \citet[see][and references therein]{jemodelshort}, power in their wavelet signatures often shifts from one to the other of the model traces.  As described in Table~\ref{orbelems}, configuration (a) was active from 2006 to 2010 and again from 2014 to 2018, while configuration (b) was active from 2002 to 2006 and again from 2010 to 2014.  The dates at which the images were taken are listed in \Fig{}~\ref{RingsresRadialScales}. 

Four of the second-order spiral bending waves shown in \Fig{}~\ref{JE2} -- namely 4:2, 8:6, 10:8, and 12:10 -- are adjacent to the first-order Janus/Epimetheus SDWs (2:1, 4:3, 5:4, and 6:5), which are among the strongest resonances in the rings \citep{Rehnberg16}.  Another three SBWs -- namely 7:5, 9:7, and 13:11 -- are adjacent to the second-order Janus/Epimetheus SDWs with the same labels, which were first studied by \citet{jemodelshort}.  The 5:2~SBW occurs in the mid-B~ring, where unexplained structure obscures the accompanying SDW.  The 14:12~SBW, which is discerned only with difficulty among the many waves due to Atlas and other moons near the very outer edge of the A~ring, is adjacent to the 7:6 resonance that governs the outer edge itself \citep{ElMoutamid16}.  

Third-order waves driven by Janus and Epimetheus are discernible only in the middle and outer parts of the A~ring, where the torques are larger (\Fig{}~\ref{JE3}).  Both SBWs and SDWs are seen for 13:10 and for 17:14, while the SDW is also seen for 14:11.  The third-order SBWs that ought to be adjacent to second-order SBWs (such as 12:9 and 15:12, respectively adjacent to 8:6 and 10:8) are generally not discernible in our data set (see \Fig{s}~\ref{JE2} and~\ref{UnexA}). 

\begin{figure}[!t]
\begin{center}
\includegraphics[height=6.8cm,keepaspectratio=true]{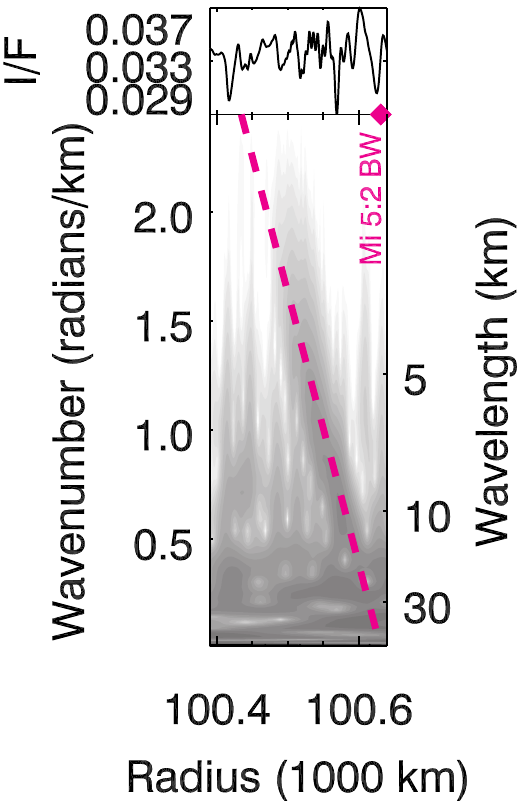}
\includegraphics[height=6.8cm,keepaspectratio=true]{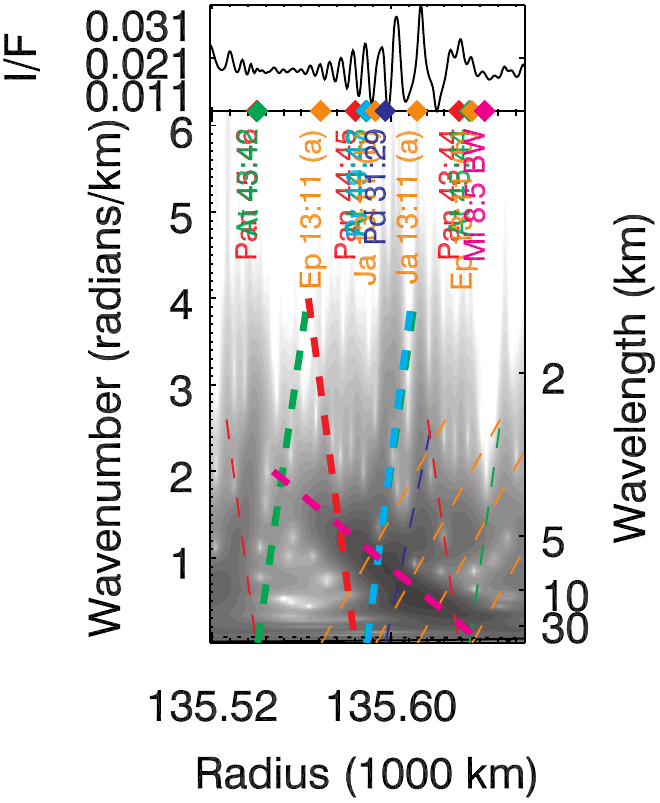}
\caption{Portions of the wavelet plots for images N1595345414 (left) and N1595337041 (right), highlighting third-order SBWs caused by the moon Mimas.  The full radial scans and wavelet plots for these images are \Fig{s}~A138 
and A110 
in Appendix~B, respectively.  In addition to the third-order Mimas SBWs shown here, the 7:4 is shown in \Fig{}~\ref{UnexA}. \label{Mi3}}
\end{center}
\end{figure}

\subsubsection{Waves driven by Mimas}

The SBWs and SDWs driven by second-order resonances with Mimas are among the strongest in the rings, and third-order SDWs from Mimas are also well known.  Third-order SBWs reported here include the 5:2 in the inner B~ring, and the 7:4 and 8:5 in the A~ring (\Fig{}~\ref{Mi3}).  Fourth-order waves from Mimas (the most likely candidate would be the Mimas~9:5 near 125,685~km) are not discernible in our data set. 

\subsubsection{Wave-like features whose cause is currently unknown \label{UnexWaves}}

The vast majority of wave-like features discernible in our data set can be matched with Lindblad resonances or vertical resonances of known moons.  The most likely explanation for those wave-like features that cannot be so explained is that they are resonating with density oscillations within the planet Saturn \citep{Rosen91b,MP93,HN13,HN14}.  The simplest explanation for the ``kronoseismology'' waves described in previous work is resonance with modes that are symmetric along lines of longitude within the planet, which would drive only SDWs.  However, recent work has uncovered what appear to be SBWs linked to $f$~modes and tesseral modes within the planet \citep{HedmanDDA16,ElMoutamid16,ElMoutamidDDA16}. 

\begin{figure}[!t]
\begin{center}
\includegraphics[height=6.8cm,keepaspectratio=true]{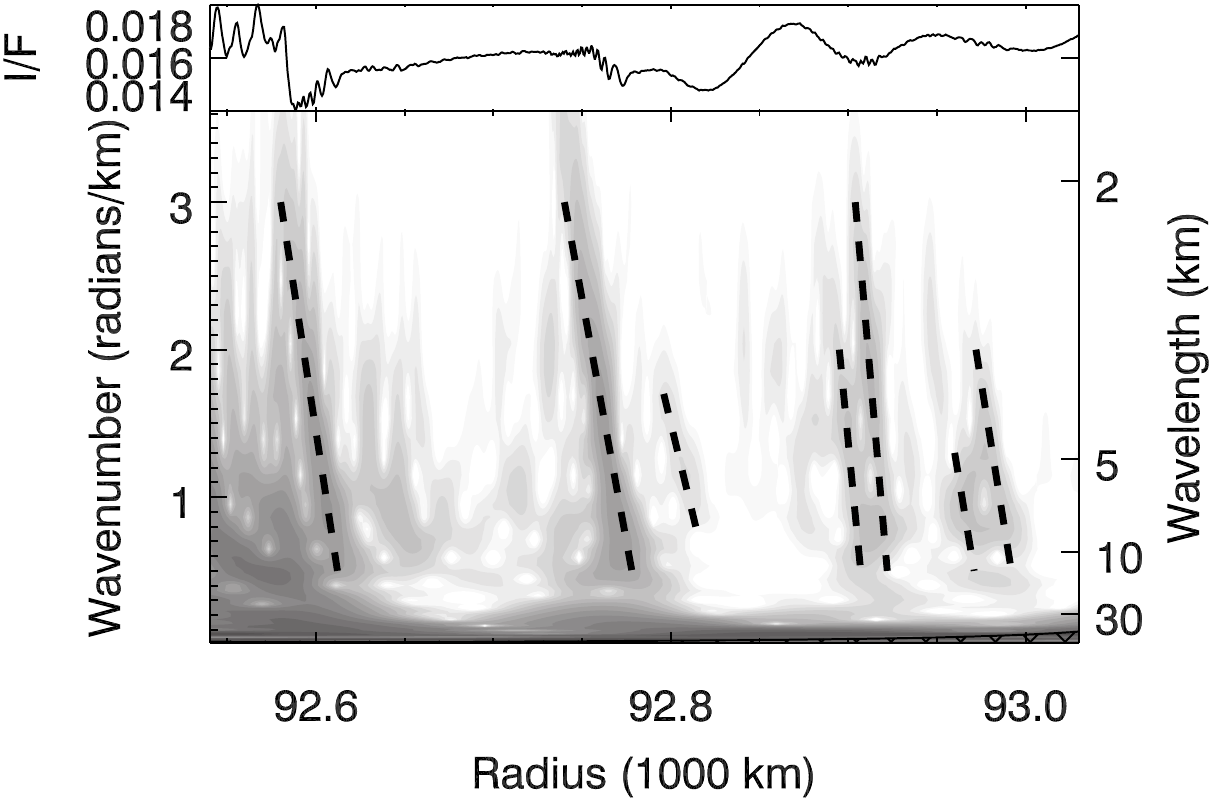}
\vspace{0.5cm}
\includegraphics[height=6.8cm,keepaspectratio=true]{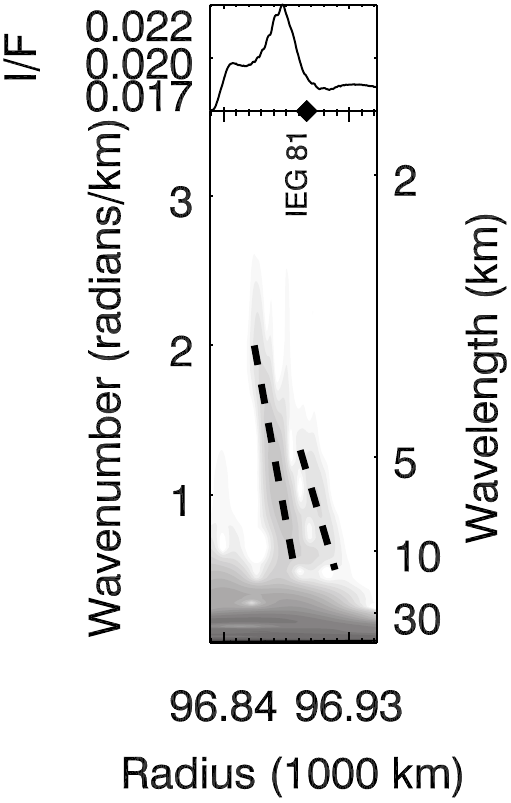}
\includegraphics[height=6.8cm,keepaspectratio=true]{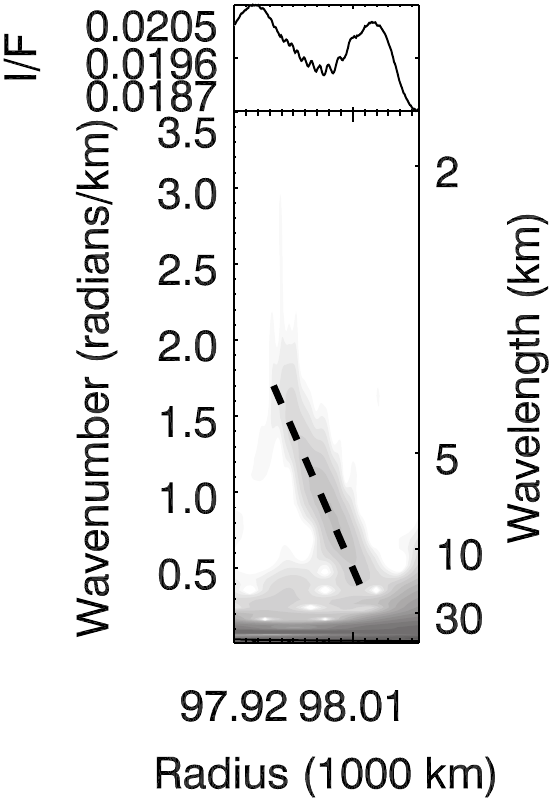}
\includegraphics[height=6.8cm,keepaspectratio=true]{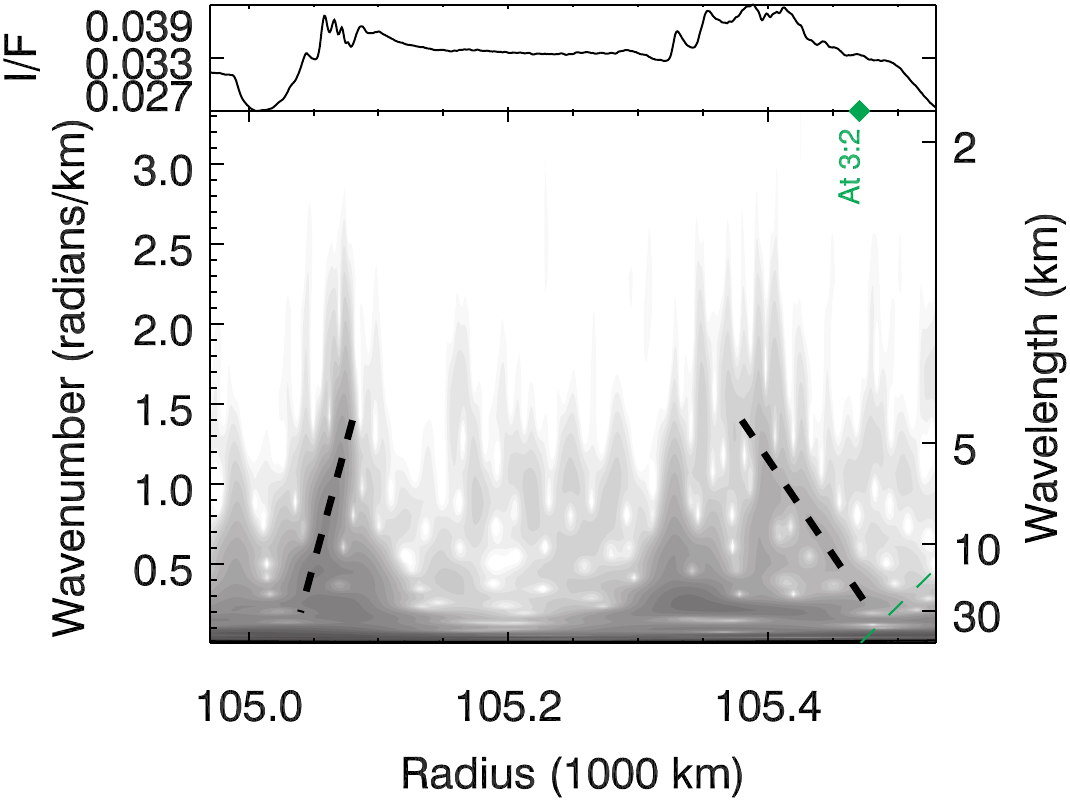}
\caption{Portions of the wavelet plots for images N1591677637 (top-left), N1591678224 (top-right), N1591678418 (bottom-left), and N1591679438 (bottom-right), highlighting unexplained features in the B~ring (black dashed lines).  The full radial scans and wavelet plots for these images are \Fig{s}~A77
, A82
, A84
, and A92 
in Appendix~B, respectively.  \label{UnexB}}
\end{center}
\end{figure}

The large majority of unexplained wave-like features in our data set appear to propagate inward, and thus are most likely to be SBWs, though they could be SDWs driven by outer Lindblad resonances (Section~\ref{OLR}).  If modes within the planet are to be considered, $f$~modes are more likely to be responsible for the former, and tesseral modes for the latter, simply based upon their frequencies.  We find 9 such inward-propagating waves in the inner B~ring, some of them in close pairs (\Fig{}~\ref{UnexB}).  In the middle B~ring, we find a paired set of an outward-propagating and an inward-propagating feature in the middle B~ring (\Fig{}~\ref{UnexB}), though they are in the wrong order to plausibly be an SBW and SDW with the same resonance label.  These features are all seen in the 077/RDHRCOMP observation, which is sensitive to vertical structure. 

\begin{figure}[!t]
\begin{center}
\includegraphics[height=6.8cm,keepaspectratio=true]{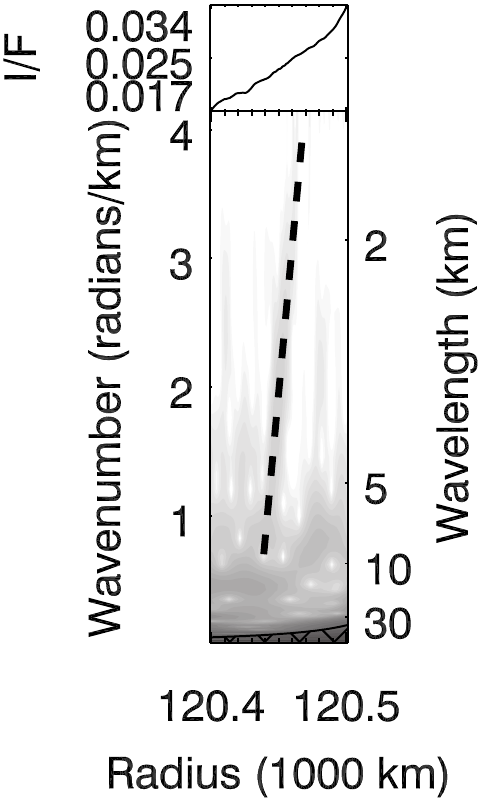} 
\includegraphics[height=6.8cm,keepaspectratio=true]{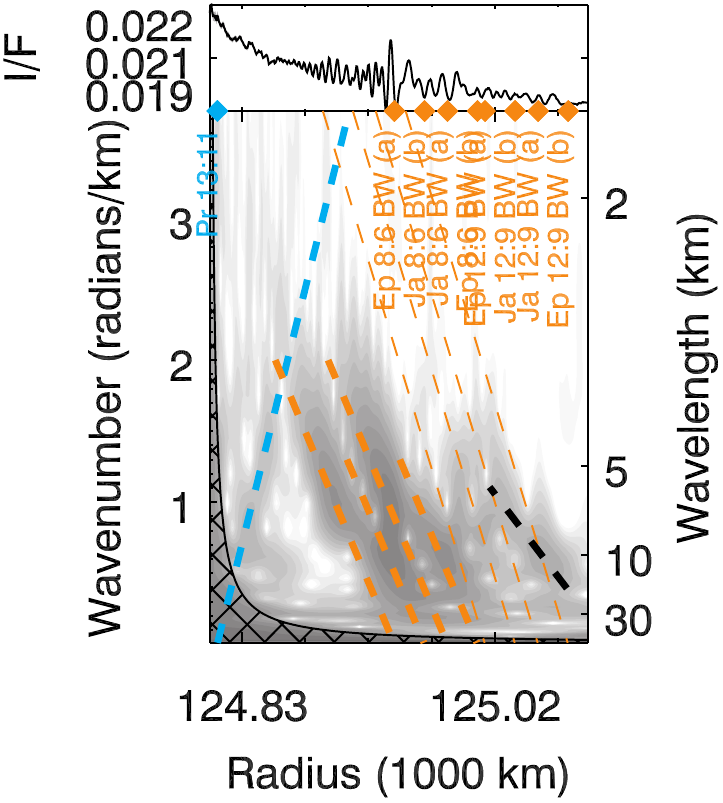} 
\includegraphics[height=6.8cm,keepaspectratio=true]{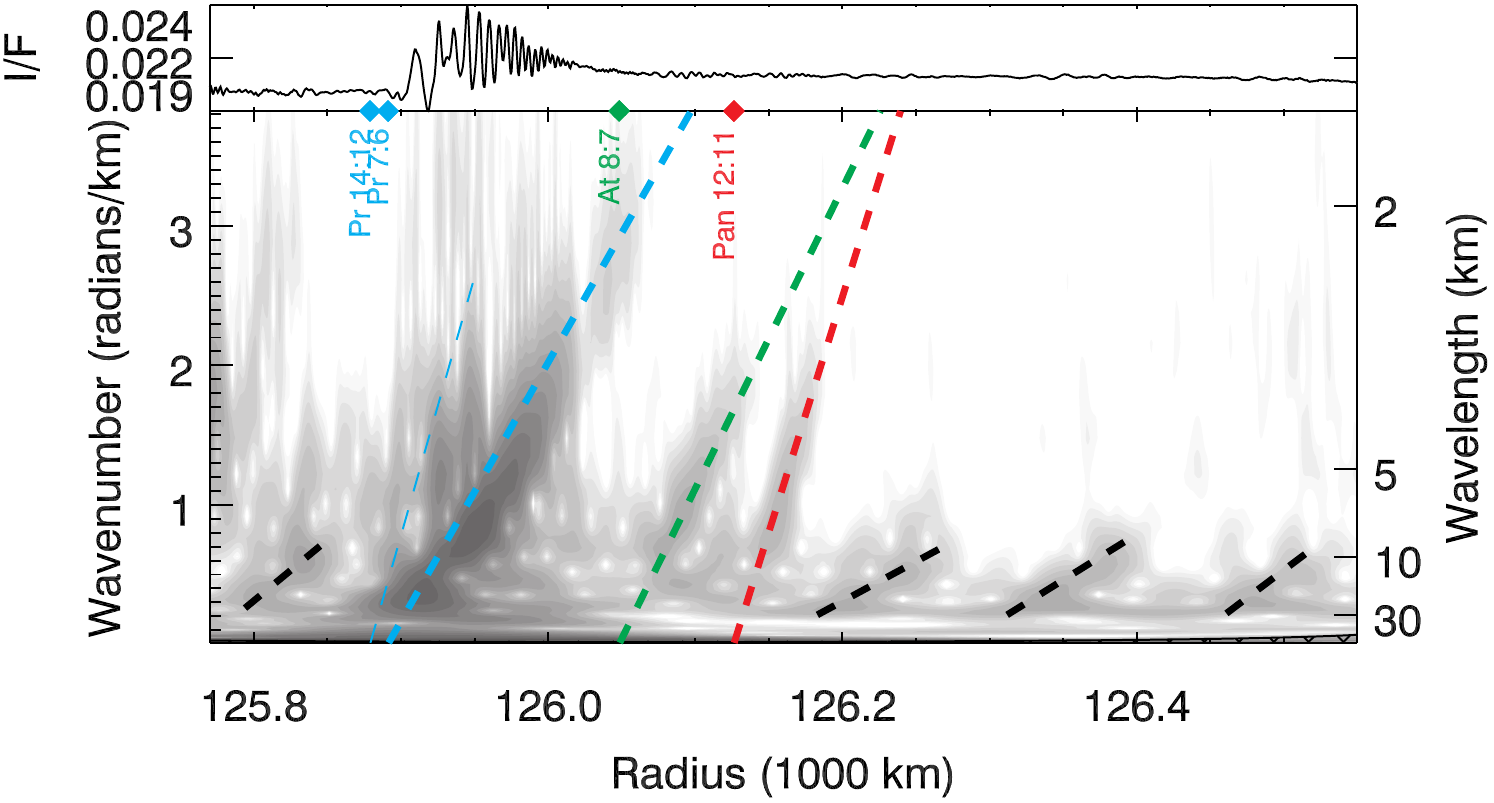}
\caption{Portions of the wavelet plots for images N1560311811 (top-left), N1595340423 (top-right), and N1595340262 (bottom), highlighting unexplained features in the Cassini Division and A~ring (black dashed lines).  The full radial scans and wavelet plots for these images are \Fig{s}~A27
, A131
, and A130 
in Appendix~B, respectively.  See continuation of this figure on the next page. \label{UnexA}}
\end{center}
\end{figure}
\renewcommand{\thefigure}{\arabic{figure} (cont'd)}
\setcounter{figure}{13}
\begin{figure}[!t]
\begin{center}
\includegraphics[height=6.8cm,keepaspectratio=true]{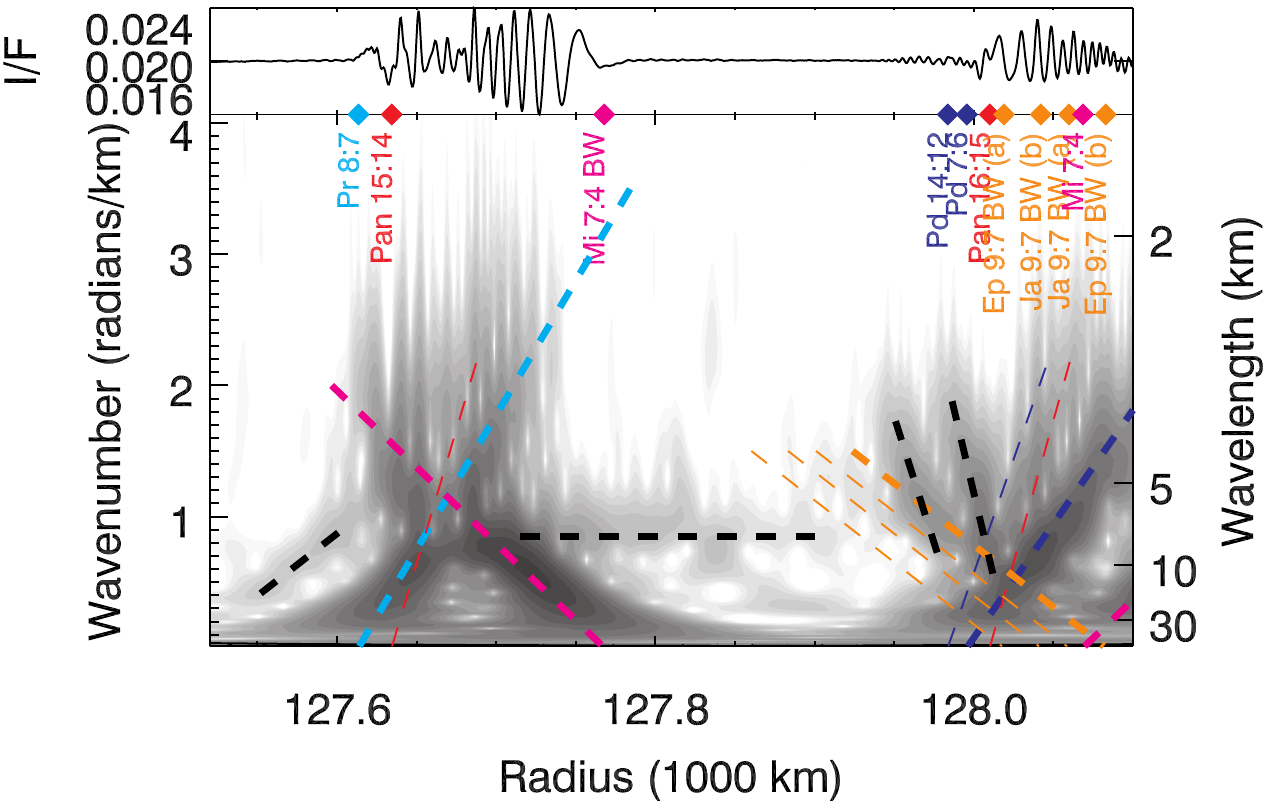}
\includegraphics[height=6.8cm,keepaspectratio=true]{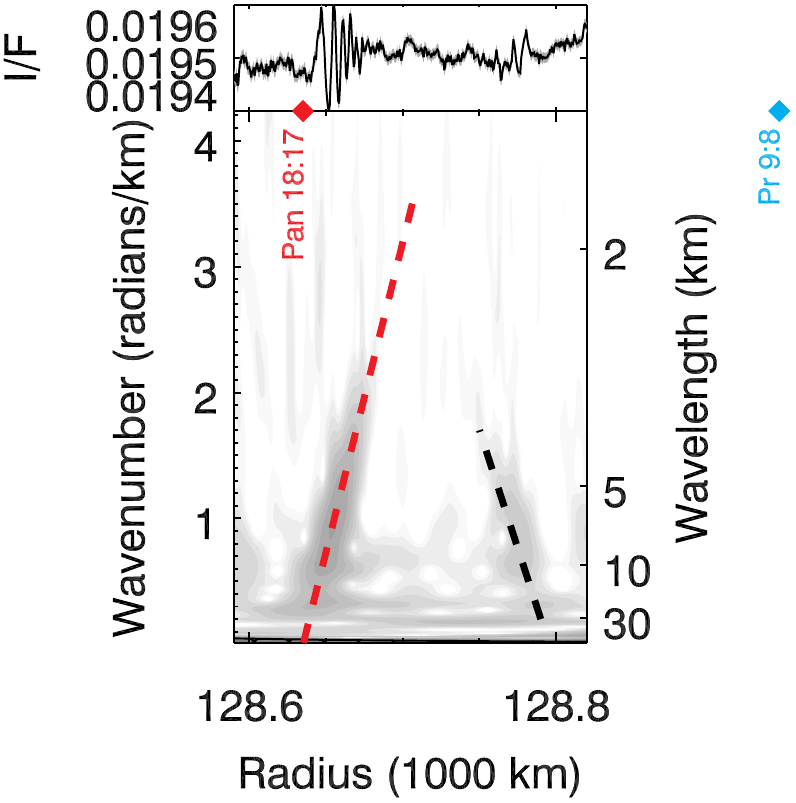}
\includegraphics[height=6.8cm,keepaspectratio=true]{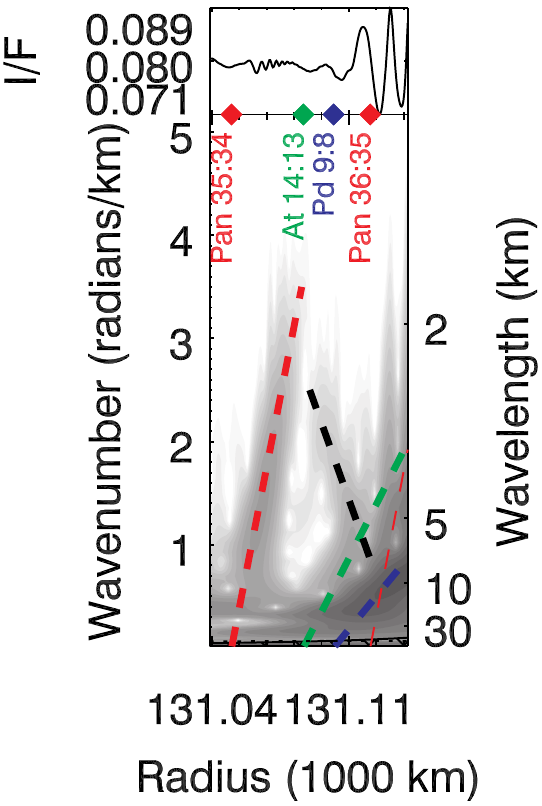}
\caption{Portions of the wavelet plots for images N1595339779 (top), N1595339457 (bottom-left), and N1560310339 (bottom-right), highlighting unexplained features in the A~ring (black dashed lines).  The full radial scans and wavelet plots for these images are \Fig{s}~A127
, A125
, and A15 
in Appendix~B, respectively.  See continuation of this figure on the next page.}
\end{center}
\end{figure}
\setcounter{figure}{13}
\begin{figure}[!t]
\begin{center}
\includegraphics[height=6.8cm,keepaspectratio=true]{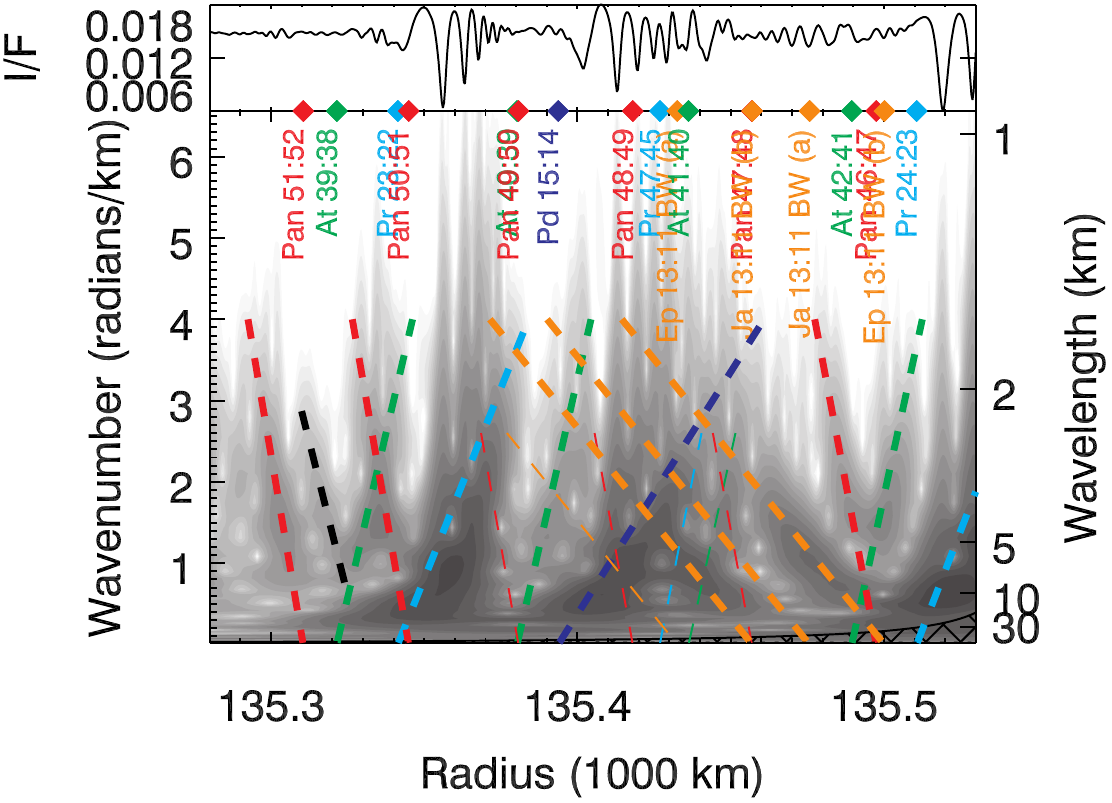}
\caption{A portion of the wavelet plot for image N1654250545, highlighting unexplained features in the A~ring (black dashed lines).  The full radial scan and wavelet plot for this image is \Fig{}~A140 
in Appendix~B.}
\end{center}
\end{figure}
\renewcommand{\thefigure}{\arabic{figure}}

At least 6 more inward-propagating wave-like features of similar morphology are seen in the A~ring (\Fig{}~\ref{UnexA}).  Another 5 unexplained features in the A~ring are inward-propagating; these all occur between 125,800 and 127,600~km and may be part of a rhythmic pattern (\Fig{}~\ref{UnexA}).  We also see an outward-propagating feature in the Cassini Division around 120,450~km.  Finally, a very unusual feature with a constant wavelength appears around 127,800~km (\Fig{}~\ref{UnexA}).  The only constant-wavelength features thought to exist in the rings are viscous overstability \citep{RL13}, but this location is not one that has previously been thought to be prone to VO. 

\begin{figure}[!t]
\begin{center}
\includegraphics[height=6.8cm,keepaspectratio=true]{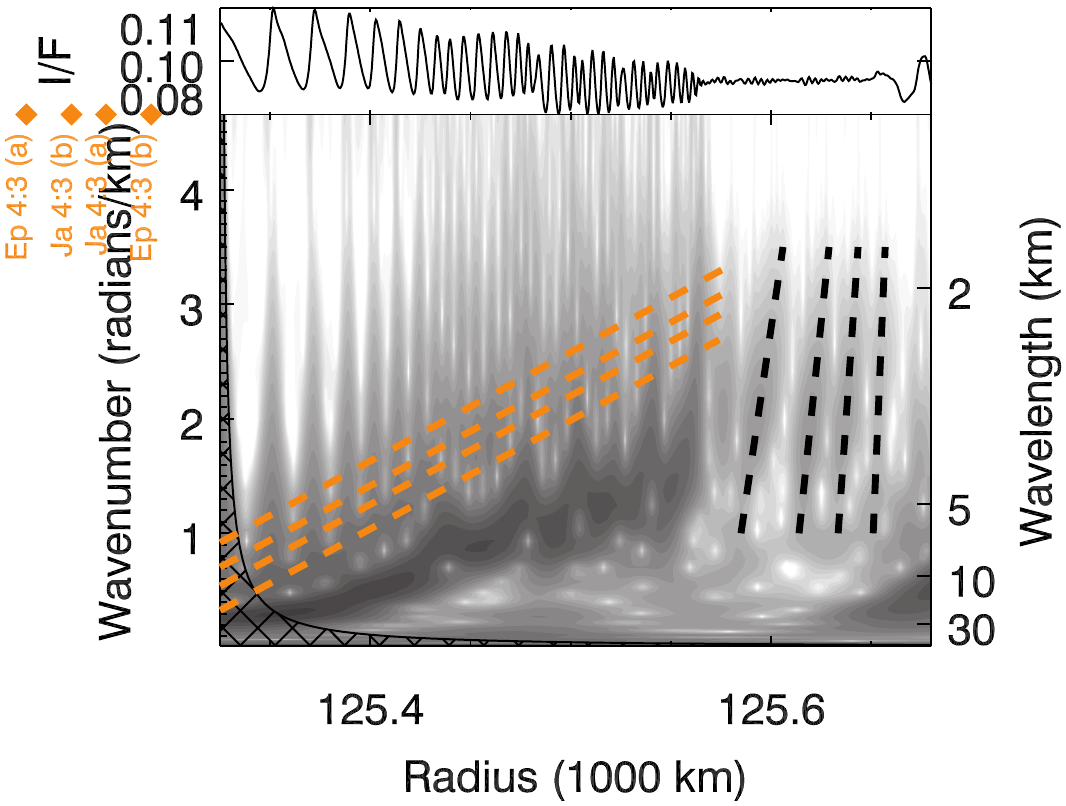}
\caption{A portion of the wavelet plot for image N1560311082, highlighting unexplained features downstream of the Janus/Epimetheus 4:3~SDW (black dashed lines).  The full radial scan and wavelet plot for this image is \Fig{}~A21 
in Appendix~B.  \label{UnexJE}}
\end{center}
\end{figure}

A final set of unexplained features are immediately downstream of the Janus/Epimetheus~4:3 SDW (\Fig{}~\ref{UnexJE}).  This is one of the strongest waves in the rings, and contains shifting irregular structure due to the 4-year orbit-swap frequency of Janus and Epimetheus.  The details of the resulting wave structure are not entirely understood \citep{jemodelshort,Rehnberg16}.  We suspect that the features highlighted in \Fig{}~\ref{UnexJE} may be remants of the Janus/Epimetheus~4:3 SDW that have evolved to appear like small waves of their own, but this is not certain. 

\subsection{Radial surface density profile \label{SurfDens}}

\begin{figure}[!t]
\begin{center}
\includegraphics[width=10cm,keepaspectratio=true]{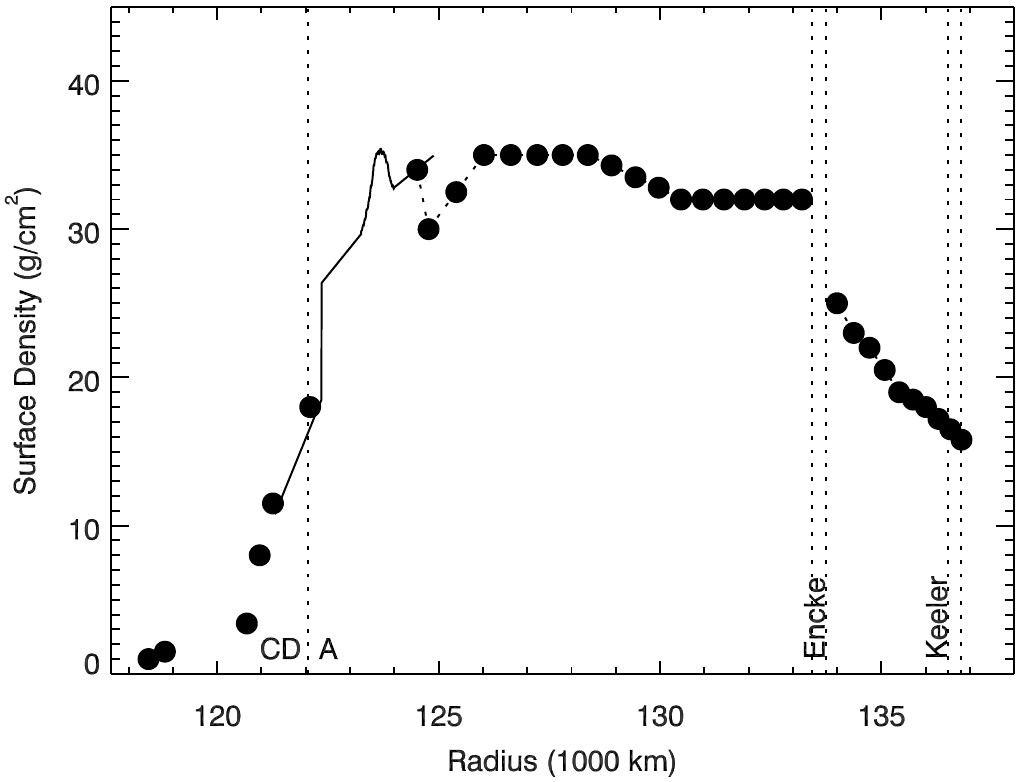}
\caption{Surface density profile for the A~ring and Cassini Division.  The filled circles are derived from the wave model fits to wavelet signatures in this work.  The filled circles in the Cassini Division (``CD'') correspond to individual SDWs and SBWs, while the filled circles in the A~ring correspond to regions fit collectively.  The solid line corresponds to the Iapetus~$-1$:0 SBW \citep{Iapetuswave13}. 
\label{SigmaRef3}}
\end{center}
\end{figure}

\begin{figure}[!t]
\begin{center}
\includegraphics[width=10cm,keepaspectratio=true]{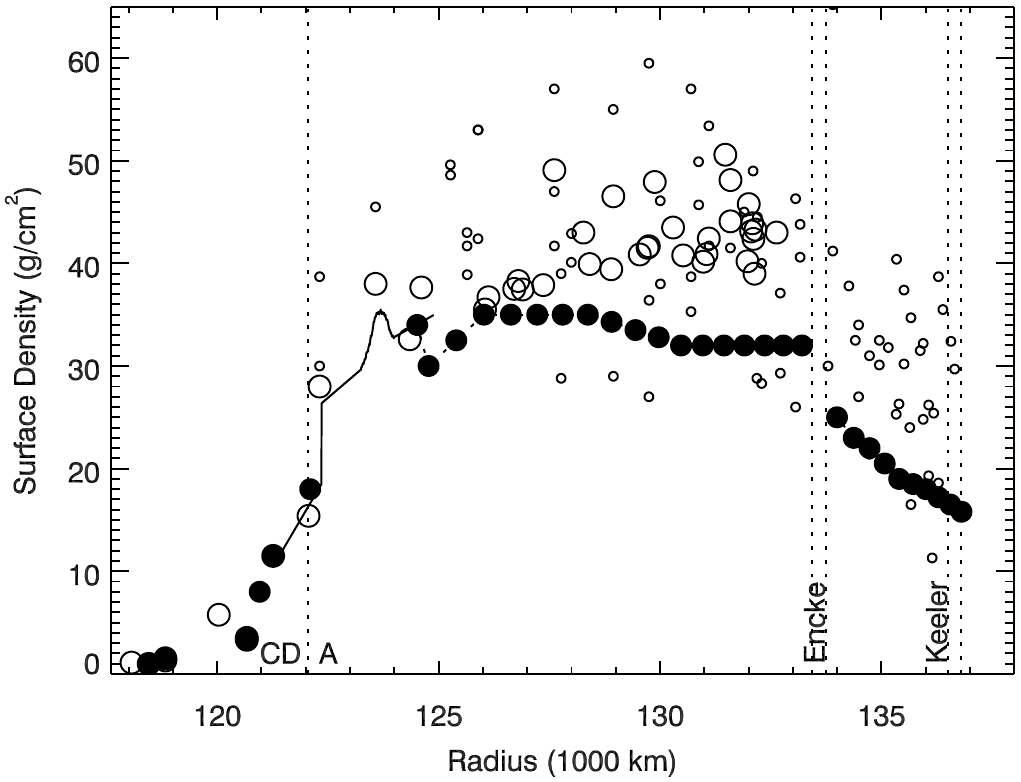}
\caption{Surface density profile comparing this work to previous work.  The filled circles and the solid line are as in \Fig{}~\ref{SigmaRef3}.  The small open circles are \Voyit{}-era data points \citep[see][for references]{soirings}, and the large open circles are data points from \citet{soirings}. 
\label{SigmaRef3vcass}}
\end{center}
\end{figure}

\begin{figure}[!t]
\begin{center}
\includegraphics[width=16cm,keepaspectratio=true]{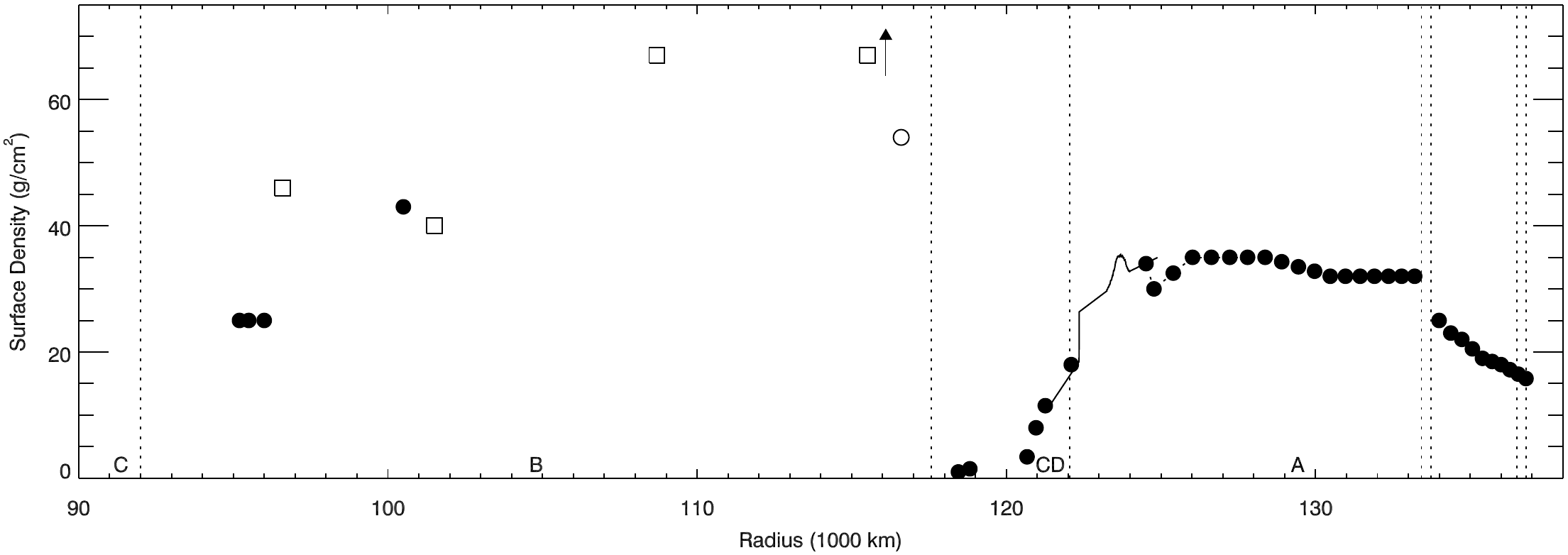}
\caption{Surface density profile including the B~ring.  The filled circles and the solid line are as in \Fig{}~\ref{SigmaRef3}.  The open squares are data points from \citet{HN16}, and the arrow indicates a surface density measurement of $\sim$130~g~cm$^{-2}$ from the same work.  The open circle is a data point from \citet{Lissauer85}, as cited by \citet{HN16}. 
\label{SigmaRef3b}}
\end{center}
\end{figure}

We present in \Fig{}~\ref{SigmaRef3} the most definitive surface density profile yet obtained for the A~ring and the Cassini Division, based on the fits displayed in the examples above and throughout Appendix~B.  We estimate the error at $\pm 1$~g~cm$^{-2}$, based on how much we could vary the background surface density $\sigma_0$ and still obtain a reasonable fit.  Although SDWs displaying non-linearity are taken into account, the fit is driven by weaker SDWs that abide by linear theory, and by SBWs.  Our assumption that $\sigma_0$ does not change very rapidly across the rings seems justified, as a single $\sigma_0$ value nearly always suffices for an entire image and the profile is quite smooth in most places.  On the other hand, some locations do display substantial trends in $\sigma_0$, as we now discuss. 

There is a sharp discontinuity in the profile at the Encke Gap.  On the inner edge of the gap, $\sigma_0 = 32$~g~cm$^{-2}$ with that value holding steady for several thousand~km inward.  On the outer edge of the gap, $\sigma_0 = 25$~g~cm$^{-2}$ (an immediate drop of 7~g~cm$^{-2}$) with the value dropping fairly steeply with ring radius down to $\sigma_0 = 16$~g~cm$^{-2}$ at the A~ring outer edge.  The region outside the Encke Gap has not previously had surface densities carefully obtained, largely because of the large number of waves and the small number of wavecrests for each wave; however, \Voyit{}-era measurements did indicate that $\sigma_0$ outward of the Encke Gap was substantially lower than inward of the gap. 

A second surprisingly sharp feature in the surface density profile occurs around 125,000~km, in the inner A~ring.  Although the Iapetus $-1$:0~SBW profile comes in around $\sigma_0 \sim 34$~g~cm$^{-2}$ at 124,500~km, the waves between that radius and 125,000~km (namely Pan~10:9, Prometheus~13:11, and the Janus/Epimetheus 8:6 SBW; see, e.g., \Fig{}~A22
) clearly fit to a value closer to $\sigma_0 \sim 30$~g~cm$^{-2}$.  However, at 126,000~km, $\sigma_0$ has risen back to $\sim$35~g~cm$^{-2}$ and holds steady with only a slight decrease over the eight thousand~km outward to the Encke Gap. 

The surface density values reported here are substantially lower than those reported previously, including by \citet{soirings}, especially in the middle A~ring (\Fig{}~\ref{SigmaRef3vcass}).  As discussed in Section~\ref{SurfDensEval} and \Fig{}~\ref{FitExample}, we believe this is due to better theoretical background and method for the current work.  

We extend our surface density profile to the B~ring in \Fig{}~\ref{SigmaRef3b}.  Information is much more sketchy for this ring region, which comprises the large majority of the ring system's mass and is pervaded almost entirely by poorly understood structure with high spatial frequencies, which makes it difficult to obtain surface densities even for the few spiral waves that occur in the region.  \citet{HN16} recently devised a novel method for combining the signatures of many VIMS occultations to increase signal-to-noise and obtained surface density values from a handful of waves.  Here we add three measurements of weak waves in the inner B~ring around 95,500~km, where the high-frequency structure is not present, as well as a clear detection of the Mimas 5:2~SBW at 100,500~km (see \Fig{}~\ref{Mi3}).  We obtain a surprisingly small $\sigma_0 \sim$ 25~g~cm$^{-2}$ around 95,500~km.  Our value of $\sigma_0 \sim$ 43~g~cm$^{-2}$ around 100,500~km corroborates nearby \citeauthor{HN16} measurements.  These numbers accentuate the inference of \citeauthor{HN16} that the total B~ring mass may be smaller than has sometimes been supposed, which increases the difficulty of maintaining the rings' observed pristine water ice composition in the face of eons of infalling meteoroid pollution, which may require the rings to be much younger than Saturn.  A definitive measurement of the ring system's mass should soon be forthcoming from direct gravity measurements as part of the Cassini Grand Finale. 

\subsection{An atlas of resonances \label{AtlasRes}}

\begin{figure}[!t]
\begin{center}
\includegraphics[width=16cm,keepaspectratio=true]{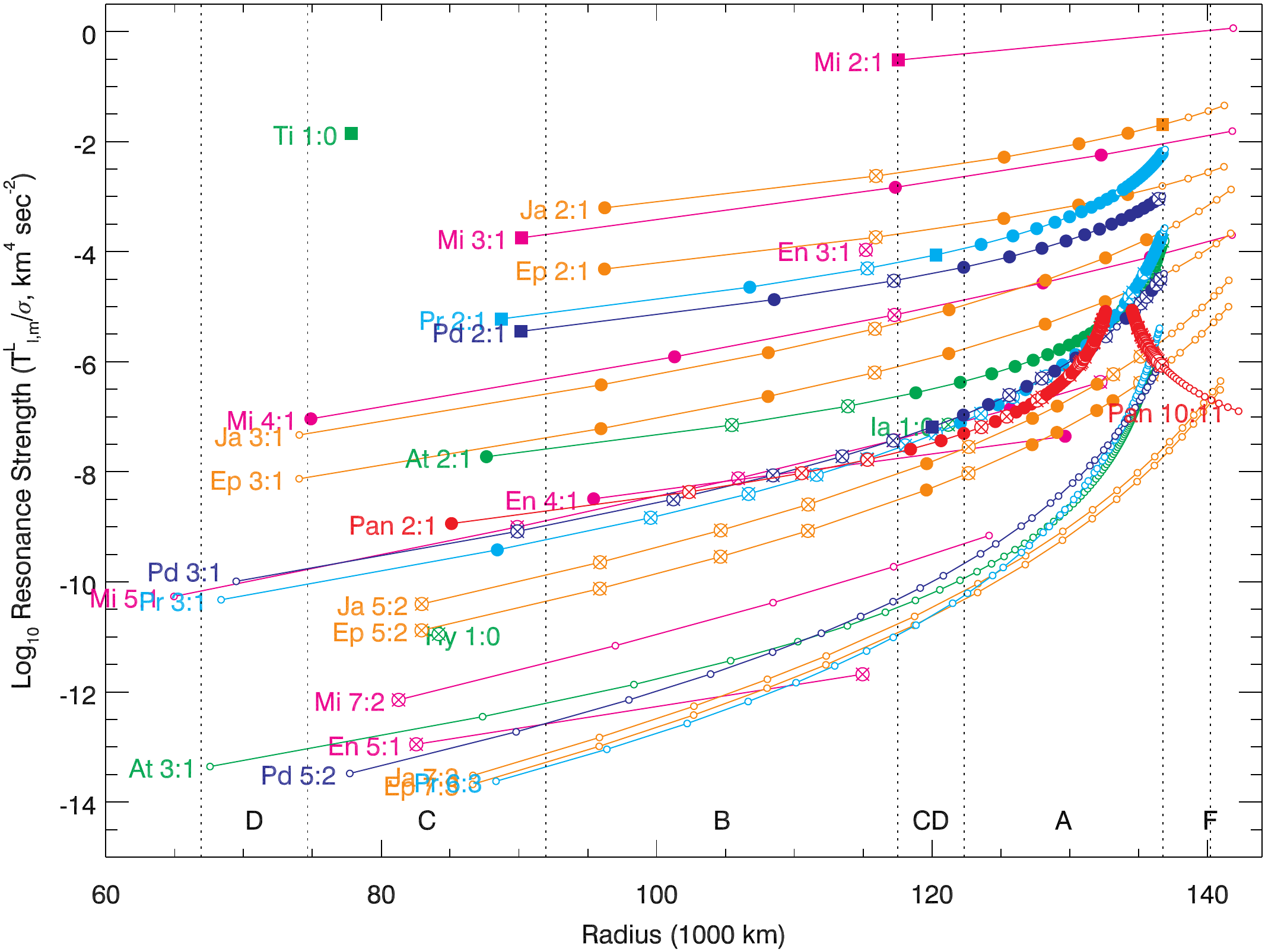}
\caption{Resonance atlas for SDWs for all of Saturn's rings, with resonances separated along the $y$-axis according to their strength.  Filled circles mean the wave is observed in our data set.  Filled squares mean the resonance is associated with an edge or a ringlet.  Open circles with an ``X'' inside mean the wave is not observed, often because a stronger wave or other structure occupies the same location.  Small open circles mean the resonance falls within a gap or outside the rings.  Resonances of the same order from the same moon are connected by lines, which are color-coded by the moon.  Only the end resonance for each group has its label printed on the plot; to move along the line, add the same number to each of the two numbers.  For example, the line beginning ``Mi~3:1'' has two other symbols, which indicate Mi~4:2 and Mi~5:3. 
\label{RestorquePlot}}
\end{center}
\end{figure}

\begin{figure}[!t]
\begin{center}
\includegraphics[height=9cm,keepaspectratio=true]{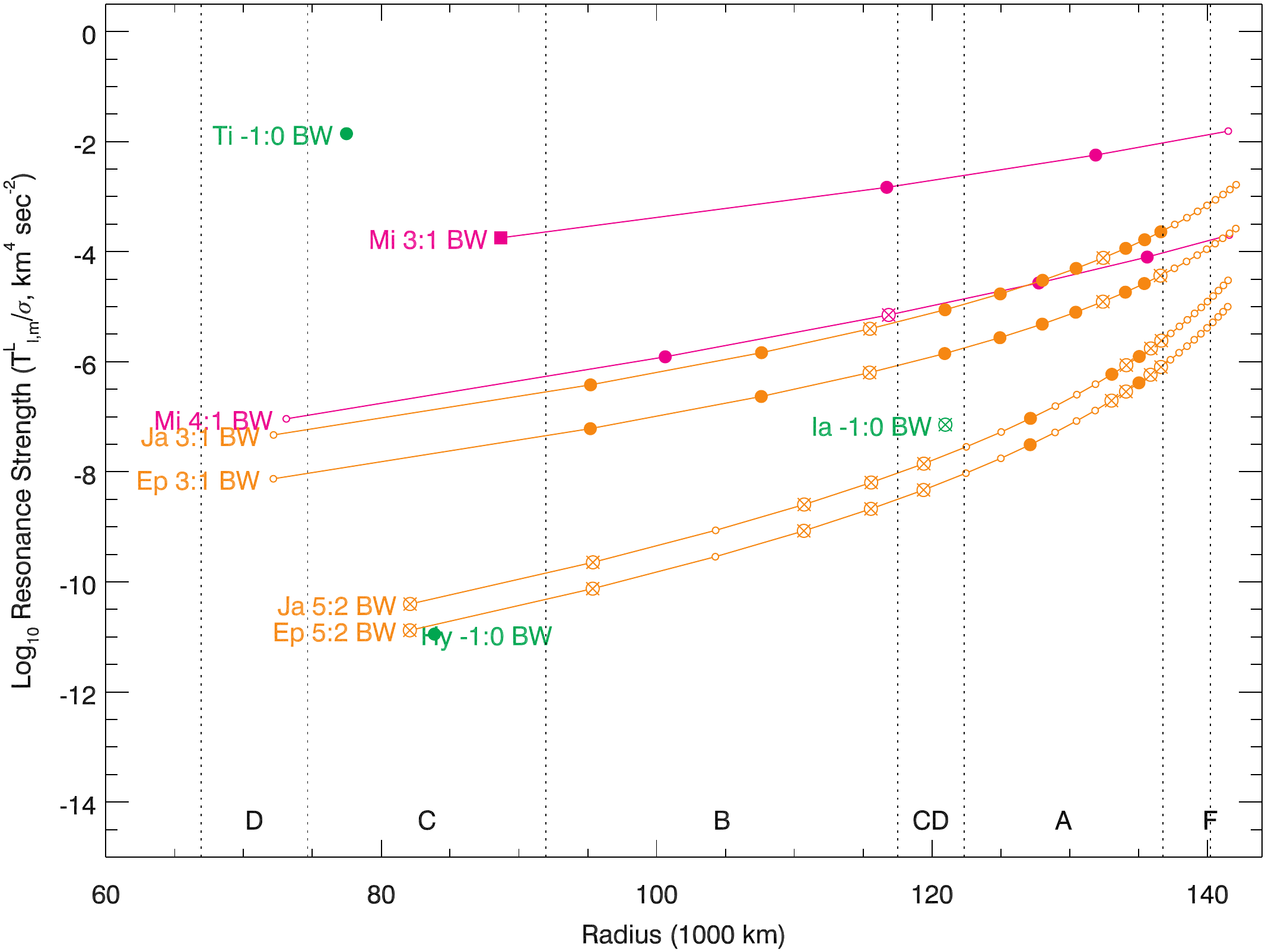}
\caption{Resonance atlas for all of Saturn's rings, like that of \Fig{}~\ref{RestorquePlot} but for SBWs instead of SDWs.  The vertical placement of the points on the plot is only approximate, due to the currently less advanced state of development for SBW theory in comparison to SDW theory (Section~\ref{VR}). 
\label{RestorquePlotSBW}}
\end{center}
\end{figure}

In \Fig{}~\ref{RestorquePlot} we update the classic plots of \citet{LC82} in light of Cassini data, showing where the relevant resonances fall in the main rings and whether or not they are observed.  This ``atlas'' can be used as a schematic for understanding the wave structure throughout the rings.  An analogous plot for SBWs is shown in \Fig{}~\ref{RestorquePlotSBW}.  In Appendix~A, \Fig{s}~A1 
through~A6 
zoom in on \Fig{}~\ref{RestorquePlot} so that each symbol can be seen clearly. 

\subsection{Towards inferring a minimum forcing for ring response}

The weakest SDWs shown in \Fig{}~\ref{RestorquePlot} have torques around $T_{m,k}/\sigma_0 \sim 10^{-8.5}$~km$^4$~sec$^{-2}$.  These include Pan~2:1 in the C~ring, Enceladus~4:1 in the inner B~ring, and Janus/Epimetheus~10:7 in the Cassini Division.  Conversely, some resonances such as Mimas~9:5 do not appear to be overwritten by stronger structure, but simply are not seen in our data set.  Observations of this type have the potential to set limits on the responsiveness of the rings to forcing, whether in terms of frequency or amplitude, which could be a new method for inferring the properties of ring particles.  At this time, we do not claim that this particular result is anything more than suggestive. 

By far the weakest bending wave in our data set is Hyperion~$-1$:0 in the C~ring.  The Janus/Epimetheus 13:10~SBW in the A~ring is only slightly weaker than the Janus/Epimetheus 4:2~SBW in the inner B~ring. 

\subsection{Towards an integrated map of Saturn's rings \label{IntegMap}}

We have made substantial progress in this work towards the long-term project of labeling all identifiable features in Saturn's rings, though much remains to be done.  The rings are both dynamic and static in an entrancing way, with a ``landscape'' defined by surface density variations and by waves, even as the waves are constantly moving spirals.  More compact features such as propellers \citep{Propellers06,Giantprops10} and impact ejecta clouds \citep{Impactclouds13} can be thought of as moving across this ``landscape.'' 

As we model and understand more details of the ring's structure, synthesized maps and interactive tools can be created to better share this knowledge with the public.  We look forward to sharing in the task of developing such products in the future. 

\vspace{1cm}
\noindent \textbf{Acknowledgements} \\We thank Joe Burns, Matt Hedman, Phil Nicholson, Mitch Gordon, Mike Evans, Maryame El~Moutamid, Radwan Tajeddine, and Carolyn Porco for helpful conversations.  We thank the Cassini Project and the Cassini Imaging Team.  The participation of B.E.H. in this project was made possible by the NSF Research Experiences for Undergraduates program, administered by the Department of Astronomy at Cornell University.  M.S.T. acknowledges funding from  the NASA Cassini Data Analysis and Participating Scientists program (NNX10AG67G, NNX13AG16G, and NNX16AH91G) and from the Cassini Project. 

\vspace{1cm}
\noindent \textbf{References}
\bibliographystyle{apalike}
\bibliography{bibliography}

\end{document}